\documentclass[a4paper,12pt]{article}

\usepackage{lmodern}            % readable font in pdf
\usepackage[T1]{fontenc}        % output special characters
\usepackage[utf8]{inputenc}     % input special characters
\usepackage{textcomp}           % more symbols
\usepackage{a4wide}             % more efficient usage of paper space

\usepackage{amsmath}            % better formulas
\usepackage{amssymb}            % math symbols
\usepackage{amsfonts}           % math fonts
\usepackage{units}              % command for units
\usepackage{slashed}            % slash symbol

\usepackage{caption}            % modify float captions
\usepackage{subfig}             % more than one figure in a float
\usepackage{cite}               % improofed citation handling

\usepackage{graphicx}           % use graphics
\usepackage{psfrag}             % latex font in plots
\usepackage[dvipsnames]{xcolor} % colors for plots
\usepackage{tikz}               % flowchart

\usepackage{booktabs}           % nicer tables
\usepackage{multirow}           % multirows for tables
\usepackage{bigstrut}           % optional extra space in tables

\usepackage{hyperref}           % pdf eigenschaften und links

\hypersetup{
    pdftitle={Quasi-stable Neutralinos at the LHC} % title
   ,pdfauthor={S. Bobrovskyi, W. Buchmüller, J. Hajer and J. Schmidt} % author
   ,pdfsubject={}                        % subject of the document
   ,pdfcreator={}                        % creator of the document
   ,pdfproducer={}                       % producer of the document
   ,pdfkeywords={R-parity violation, Susy, LHC, CMS} % list of keywords
   ,colorlinks=true                      % false: boxed links; true: colored links
   ,linkcolor=black                      % color of internal links
   ,citecolor=black                      % color of links to bibliography
   ,filecolor=black                      % color of file links
   ,urlcolor=black                       % color of external links
   ,pdfdisplaydoctitle=true              % windowtitle is identical to document title, not filename
   ,bookmarksnumbered=true               % nummeratet pdf table of contents
}

\usetikzlibrary{shapes}
\usetikzlibrary{arrows}
\usetikzlibrary{positioning}

\newcommand{\software}[1]{\texttt{\uppercase{#1}}}

\title{
\vspace{-4.5ex}
{\normalsize \raggedright
DESY 11--077\\
July 2011\\[10ex]
}
\textbf{Quasi-stable neutralinos at the LHC}
\vspace{2ex}
}

\author{S.~Bobrovskyi, W.~Buchmüller, J.~Hajer and J.~Schmidt\\[1ex]
{\normalsize\textit{Deutsches Elektronen-Synchrotron DESY, Hamburg, Germany}}
\vspace{3ex}
}
\date{}

\begin{document}

\maketitle

\thispagestyle{empty}

\begin{abstract}
\noindent
We study supersymmetric extensions of the Standard Model with small R-parity and lepton number violating couplings which are naturally consistent with primordial nucleosynthesis,
thermal leptogenesis and gravitino dark matter.
We consider supergravity models where the gravitino is the lightest superparticle followed by a bino-like next-to-lightest superparticle (NLSP).
Extending previous work we investigate in detail the sensitivity of LHC experiments to the R-parity breaking parameter~$\zeta$ for various gluino and squark masses.
We perform a simulation of signal and background events for the generic detector \software{DELPHES} for which we implement the finite NLSP decay length.
We find that for gluino and squark masses accessible at the LHC, values of $\zeta$ can be probed which are one to two orders of magnitude smaller than the present upper bound obtained from astrophysics and cosmology.
\end{abstract}

\newpage

\section{Introduction}

Supersymmetric extensions of the Standard Model with broken R-parity have a rich
phenomenology \cite{Hall:1983id,Ross:1984yg,Ellis:1984gi,Allanach:2003eb,
Barbier:2004ez}. In most models rather large R-parity violating couplings are
considered, which lead to prompt decays of the lightest superparticle in the
detector. In models where small R-parity violating interactions generate neutrino
masses, macroscopic decay lengths up to \unit[1]{mm} are obtained
\cite{deCampos:2007bn}. In the case of gauge mediated supersymmetry breaking,
R-parity violating decays then compete with R-parity conserving decays where the
final state contains a gravitino \cite{Hirsch:2005ag}.

If the gravitino is the lightest superparticle, its decays are doubly suppressed
by the Planck mass and the small R-parity breaking parameter. Hence, its lifetime
typically exceeds the age of the universe by many orders of magnitude, so that it
remains a viable dark matter candidate \cite{Takayama:2000uz}. In the case of
very small R-parity breaking couplings, as they occur if R-parity
is spontaneously broken at the grand unification scale, one obtains a consistent
cosmology including primordial nucleosynthesis, thermal leptogenesis and gravitino
dark matter \cite{Buchmuller:2007ui}. In the following we shall study the
collider phenomenology of these models extending our previous work
\cite{Bobrovskyi:2010ps}.

Decaying gravitino dark matter leads to a diffuse gamma-ray flux
\cite{Takayama:2000uz,Buchmuller:2007ui,Lola:2007rw}. For a bino-like NLSP the
matrix elements for gravitino decay and NLSP decay are directly related
\cite{Bobrovskyi:2010ps}. Together with the lower bound on the gravitino lifetime,
and the corresponding upper bound on the R-parity breaking parameter $\zeta$,
which is derived from the diffuse gamma-ray flux observed by the Fermi-LAT
collaboration \cite{Abdo:2010nc,Abdo:2010nz}, one obtains a lower bound on the
NLSP decay length.
The estimates in \cite{Bobrovskyi:2010ps} led to $\zeta \lesssim 3\times 10^{-8}$,
and in the more detailed analysis of gamma-ray lines in \cite{Vertongen:2011mu}
a slightly stronger bound was found, $\zeta \lesssim 2\times 10^{-8}$.
For gravitino masses
$\mathcal{O}(\unit[100]{GeV})$ the upper bound becomes more stringent:
$\zeta \sim 10^{-9}$ \cite{Bobrovskyi:2010ps}. This range in the parameter $\zeta$
corresponds to NLSP decay lengths varying from $\mathcal{O}(\unit[50]{cm})$ to
$\mathcal{O}(\unit[500]{m})$.

Large macroscopic decay lengths are of great help in the search for decaying NLSPs.
This remains true if the decay length is larger than the size of the detector
since a sizeable fraction of NLSPs may still decay inside the detector. This has
been studied for neutral \cite{Ishiwata:2008tp} as well as charged \cite{Asai:2009ka}
NLSPs. Neutralino decay lengths varying from \unit[0.1]{mm} to \unit[100]{m}
also arise in models with generalized gauge mediated supersymmetry breaking
\cite{Meade:2010ji}. Alternatively, charged \cite{Desch:2010gi}
and neutral \cite{Bomark:2011ye} NLSP decays have been studied for models
where the decay lengths are so small that no displaced vertices are observed and
R-parity breaking Yukawa couplings determine the hierarchy of decay channels. In
this case multi-lepton events, and their flavour structure, are of crucial
importance.

The subject of this paper is a quantitative analysis of neutralino NLSP decays at
the LHC in the case of very small R-parity breaking,
$\zeta \lesssim 3\times 10^{-8}$. The goal is the determination
of the sensitivity in $\zeta$ for varying gluino and squark masses. To achieve this
we perform a simulation for the generic detector \software{DELPHES} \cite{Ovyn:2009tx} of signal
and background events. We focus on events with a clean signature: cascade processes
with jets where one of the produced neutralino NLSPs decays into $Z$-boson and
neutrino, with the subsequent decay of the $Z$-boson into a muon pair. This allows
us to determine a conservative 5$\sigma$ discovery range. Finally, we
estimate the discovery reach of the LHC if all NLSP decays are taken into account.

The paper is organized as follows.
In Section~\ref{sec:r_parity_breaking} we recall the main ingredients of the considered model of bilinear R-parity breaking,
as well as the dominant NLSP decays into gauge bosons and leptons. In Section~\ref{sec:signatures_and_reach} we discuss qualitative signatures of the events:
NLSP $\beta\gamma$ distribution, $\slashed p_T$ spectrum, and the number of final state leptons.
In Section~\ref{sec:signal_and_background}, five representative benchmark points are defined
and the simulation of signal and background events is described with emphasis on the muon reconstruction.
The numerical results for the chosen NLSP decay channel $Z(\mu^+\mu^-)\nu$ are given in Section~\ref{sec:neutrsearch},
together with estimates for the discovery reach if all NLSP decays are included. We conclude with a summary in Section~\ref{sec:summary}.

\section{Connecting neutralino and gravitino decays}
\label{sec:r_parity_breaking}

\subsection{Bilinear R-parity breaking}

We consider supersymmetric extensions of the Standard Model with bilinear R-parity
breaking (cf.~\cite{Allanach:2003eb, Barbier:2004ez}) as they are obtained if the
spontaneous breaking of B$-$L, the difference of baryon and lepton number, is
related to the spontaneous breaking of R-parity \cite{Buchmuller:2007ui}.
Mass mixing terms between lepton and Higgs fields then appear in the
superpotential\footnote{Our notation for Higgs and matter superfields, scalars and
left-handed fermions reads: $H_u = (H_u, h_u)$, $l_i = (\tilde l_i, l_i)$ etc.},
\begin{equation}
\Delta W = \mu_i H_u l_i \ ,
\label{rpvw1}
\end{equation}
as well as the scalar potential induced by supersymmetry breaking,
\begin{equation}
 - \Delta \mathcal L
 = B_i H_u \tilde l_i
 + m^2_{id} \tilde l^\dagger_i H_d
 + \text{h.c.} \ .
\label{rpvv1}
\end{equation}
These mixing terms, together with the R-parity conserving superpotential
\begin{equation}
    W
  = \mu H_u H_d
  + h_{ij}^u q_i u^c_j H_u
  + h_{ij}^d d^c_i q_j H_d
  + h_{ij}^e l_i e^c_j H_d \ ,
\label{w1rp}
\end{equation}
the scalar mass terms
\begin{align}
  - \mathcal L_{\text{M}}
 =&\; m^2_u H_u^\dagger H_u
  + m^2_d H_d^\dagger H_d
  + \left( B H_u H_d + \text{h.c.} \right)
 \nonumber \\
 &+ \widetilde m^2_{li} \tilde l_i^\dagger \tilde l_i
  + \widetilde m^2_{ei} \tilde e_i^{c\dagger} \widetilde e^c_i
  + \widetilde m^2_{qi} \widetilde q_i^\dagger \widetilde q_i
  + \widetilde m^2_{ui} \widetilde u_i^{c\dagger} \widetilde u^c_i
  + \widetilde m^2_{di} \widetilde d_i^{c\dagger} \widetilde d^c_i \ ,
\label{scalarmass}
\end{align}
and the standard $\mathrm{SU(3)\times SU(2)\times U(1)_Y}$ gauge interactions
define the supersymmetric standard model with bilinear R-parity breaking.
Note that the Higgs mass terms $m_u^2$ and $m_d^2$ contain the contributions
both from the superpotential~\eqref{w1rp} and the soft supersymmetry breaking
terms. For simplicity, we have assumed flavour diagonal mass matrices
in~\eqref{scalarmass}.

As discussed in \cite{Bobrovskyi:2010ps}, it is convenient to work in a basis of
$\mathrm{SU(2)}$ doublets where the mass mixings $\mu_i$, $B_i$ and $m_{id}^2$
in Eqs.~\eqref{rpvw1} and~\eqref{rpvv1} are traded for R-parity breaking Yukawa
couplings. This can be achieved by field redefinitions: the standard rotation
of the superfields $H_d$ and $l_i$,
\begin{equation}
   H_d
 = H_d^\prime
 - \epsilon_i l^\prime_i
   \ ,\quad
   l_i
 = l^\prime_i
 + \epsilon_i H_d^\prime
   \ ,\quad
   \epsilon_i
 = \frac{\mu_i}{\mu} \ ,
\end{equation}
followed by a non-supersymmetric rotation involving all scalar
$\mathrm{SU(2)}$ doublets,
\begin{equation}
   H^\prime_d
 = H^{\prime\prime}_d
 - \epsilon^\prime_i \tilde l^{\prime\prime}_i
   \ ,\quad
   \varepsilon H^*_u
 = \varepsilon H^{\prime*}_u
 - \epsilon^{\prime\prime}_i \tilde l^{\prime\prime}_i
   \ ,\quad
   \tilde l^\prime_i
 = \tilde l^{\prime\prime}_i
 + \epsilon^\prime_i H^{\prime\prime}_d
 + \epsilon^{\prime\prime}_i \varepsilon H^{\prime*}_u
\;,
\end{equation}
where $\epsilon_i^\prime$ and $\epsilon_i^{\prime\prime}$ are functions of
$B$, $B_i$, $m_{id}^{2}$, $\widetilde m_{li}^2$, $m_u^2$ and $m_d^2$
\cite{Bobrovskyi:2010ps}.

The R-parity breaking Yukawa terms contain couplings between gauginos,
lepton doublets and Higgs doublets. After electroweak symmetry breaking,
$\langle H_u^0 \rangle = v_u$, $\langle H_d^0 \rangle = v_d$,
one obtains new mass mixings between higgsinos, gauginos and leptons\footnote{
Our notation for gauge fields and left-handed gauginos reads:
$B_{\mu}$, $b$ etc.},
\begin{align}\label{mixing}
-\Delta {\mathcal L}_M \supset \
m^e_{ij}\frac{\zeta_i}{c_\beta} e^c_j h_d
 - m_Z s_w \zeta_i^* \nu_i b
 + m_Z c_w \zeta_i^* \nu_i w^3 + \text{h.c.} \ ,
\end{align}
where we have defined
\begin{align}
\zeta_i &= \frac{\epsilon_i^\prime v_d + \epsilon_i^{\prime\prime} v_u}{v}\ , \quad
v = \sqrt{v_u^2 + v_d^2}\ , \quad \frac{v_u}{v_d} =
\tan{\beta} \equiv \frac{s_\beta}{c_\beta} \ ,
\label{zetabeta} \\
m^e_{ij} &= h_{ij}^{e}v_d\ , \quad
m_Z = \frac{\sqrt{g^2 + g^{\prime2}}v}{\sqrt{2}} \ , \quad
s_w = \frac{g^\prime}{\sqrt{g^2 + g^{\prime2}}} = \sqrt{1 - c_w^2}\ .
\end{align}
Here $g$, $g^\prime$ and $h_{ij}^{e}$ are the $\mathrm{SU(2)}$ and the $\mathrm{U(1)_Y}$
gauge couplings and the charged lepton Yukawa couplings, respectively.

The diagonal mass terms together with the mixing terms in Eq. \eqref{mixing} represent
the $7\times 7$ neutralino mass matrix of the gauginos $b$, $w^3$, the higgsinos
$h_u^0$, $h_d^0$ and the three neutrinos $\nu_i$, and also the $5\times 5$ chargino mass
matrix of gaugino, higgsino and charged leptons. Both mass matrices have to be
diagonalized to obtain the CKM-type matrix elements of neutral
($V^{(\chi,\nu)}$), charged ($V^{(\chi,e)}$) and supercurrents
($U^{(\chi,\nu)}$). For the standard supergravity mass spectrum with a bino-like
NLSP $\chi_1^0$ one obtains the R-parity breaking matrix elements
($s_{2\beta} = 2 s_\beta c_\beta$) \cite{Bobrovskyi:2010ps}:
\begin{align}
V_{1i}^{(\chi,\nu)} =& - \zeta_i \frac{m_Z s_w}{2 M_1}
\left( 1+\mathcal O \left(s_{2 \beta}\frac{m_Z^2}{\mu^2} \right) \right) \ ,
\label{meneutral} \\
V_{1i}^{(\chi,e)}=&- \zeta_i \frac{m_Z s_w}{ M_1}
\left( 1+\mathcal O \left(s_{2 \beta} \frac{m_Z^2}{\mu^2}\right)\right) \ ,
\label{mecharged} \\
U_i^{(\tilde \gamma,\nu)}=& \ \zeta_i \frac{m_Z s_w c_w
\left(M_2-M_1\right)}{M_1 M_2}
\left(1+\mathcal O \left(s_{2 \beta}\frac{m_Z^2}{\mu^2}\right)
\right) \ , \label{gravphoton}
\end{align}
where the photino matrix element is defined as
\begin{align}
U^{(\widetilde\gamma,\nu)}_i = c_w U^{(b,\nu)}_i + s_w U^{(w,\nu)}_i\ .
\end{align}
Note that the charged and
neutral current matrix elements agree up to the isospin factor
at leading order in $m_Z^2/\mu^2$,
i.e., $V^{(\chi,\nu)}_{1i\ \text{LO}} = V^{(\chi,e)}_{1i\ \text{LO}}/2$.

\subsection{Gravitino and NLSP decays}
\label{sec:neutralino_signatures}

\begin{figure}
\centering
\includegraphics[width=0.4\textwidth]{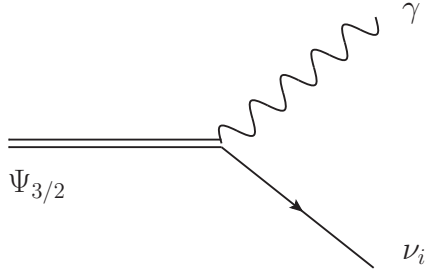}
\caption{Gravitino decay into photon and neutrino.}
\label{fig:gravitino}
\end{figure}

The partial width for gravitino decay into photon and neutrino
(cf.~Fig.~\ref{fig:gravitino}) is given by \cite{Takayama:2000uz}
\begin{align}\label{widthgravitino}
\Gamma_{\nicefrac{3}{2}}(\gamma\nu) =
\frac{1}{32\pi} \sum_i\left|U^{(\widetilde\gamma, \nu)}_i\right|^2
\frac{m_{\nicefrac{3}{2}}^3}{M_{\mathrm{P}}^2}\ .
\end{align}
Inserting the matrix element (\ref{gravphoton}) one obtains for the gravitino
lifetime to leading order in $m_Z/\mu$ \cite{Bobrovskyi:2010ps}:
\begin{align}\label{gravlife2}
\tau_{\nicefrac{3}{2}}(\gamma\nu) =
\unit[1 \times 10^{27}]{s}
\left(\frac{\zeta}{10^{-7}}\right)^{-2}
\left(\frac{m_{\chi^0_1}}{\unit[100]{GeV}}\right)^{2}
\left(\frac{m_{\nicefrac{3}{2}}}{\unit[10]{GeV}}\right)^{-3}\ ,
\end{align}
where
\begin{align}
\zeta^2 = \sum_i \zeta_i^2 \ .
\end{align}
Based on the Fermi-LAT search for dark matter decaying into two photons, upper
bounds on the R-parity breaking parameter $\zeta$ were derived:
$\zeta \lesssim 3\times 10^{-8}$
\cite{Bobrovskyi:2010ps} and $\zeta \lesssim 2\times 10^{-8}$ \cite{Vertongen:2011mu}.

\begin{figure}
%\centering
\psfrag{\chi}{\small$\chi_1^0$}
\psfrag{\nu}{\small$\nu_i$}
\psfrag{l}{\small$l^-$}
\psfrag{0}{}
\psfrag{1}{}
\psfrag{i}{}
\psfrag{Z}{\small$Z$}
\psfrag{W}{\small$W^+$}
\psfrag{+}{}
\psfrag{-}{}
\subfloat{\includegraphics[width=0.40\textwidth]{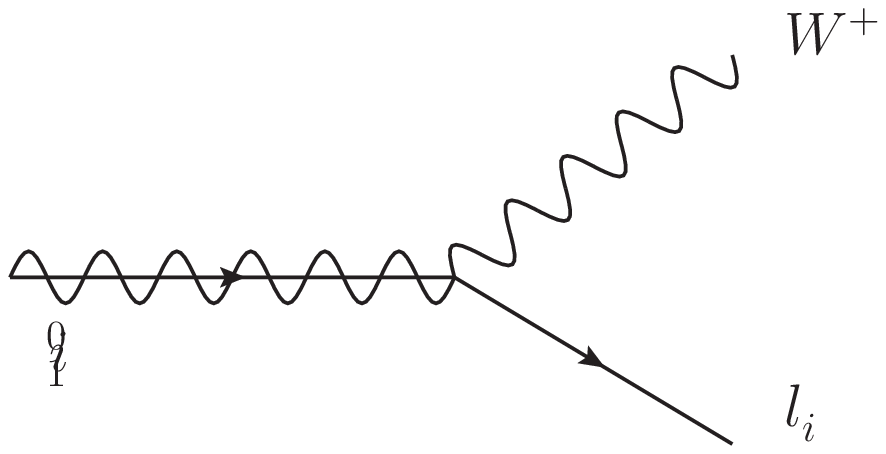}}
\hspace{0.1\textwidth}
\subfloat{\includegraphics[width=0.40\textwidth]{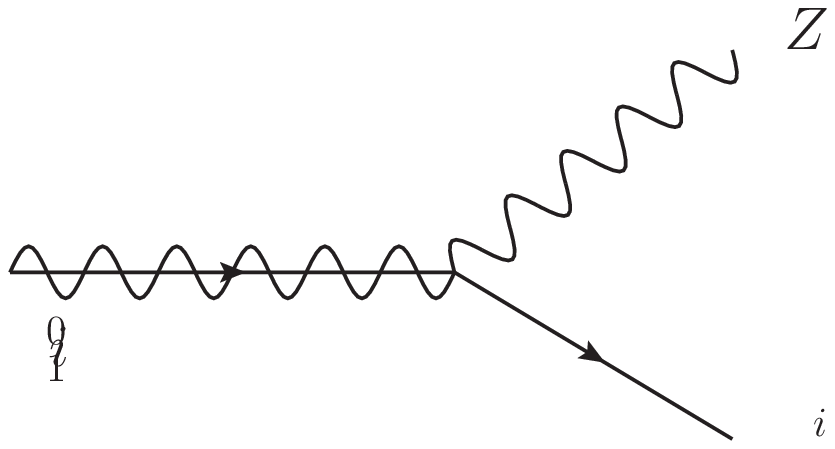}}
\caption{Neutralino decays into charged lepton and $W$-boson,
and neutrino and $Z$-boson.}
\label{fig:neutralinodecay}
\end{figure}
\begin{figure}
\centering
\psfrag{mchiinGeV}{\small$m_{\chi_1^0}$~[GeV]}
\psfrag{fzandfw}{\small$f$}
\psfrag{100}{\footnotesize100}
\psfrag{150}{\footnotesize150}
\psfrag{200}{\footnotesize200}
\psfrag{250}{\footnotesize250}
\psfrag{300}{\footnotesize300}
\psfrag{350}{\footnotesize350}
\psfrag{400}{\footnotesize400}
\psfrag{0.2}{\footnotesize\hspace{-0.3em}0.2}
\psfrag{0.4}{\footnotesize\hspace{-0.3em}0.4}
\psfrag{0.6}{\footnotesize\hspace{-0.3em}0.6}
\psfrag{0.8}{\footnotesize\hspace{-0.3em}0.8}
\psfrag{1.0}{\footnotesize\hspace{-0.3em}1.0}
\includegraphics[width=0.6\textwidth]{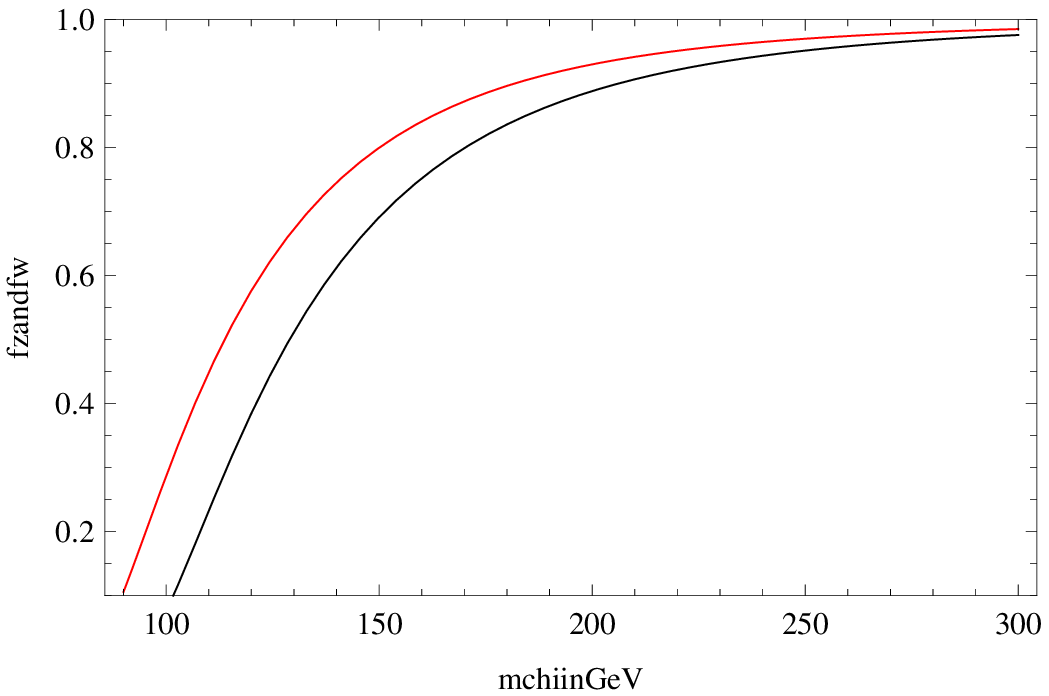}
\caption{Phase space suppression factors for neutralino decays into $W$-bosons (red) and $Z$-bosons (black) and leptons.}
\label{fig:phasespace}
\end{figure}

The parameter $\zeta$ also controls the lifetime of the bino-like NLSP
$\chi_1^0$. For a NLSP mass larger than 100~GeV decays into charged lepton
and $W$-boson or neutrino and $Z$-boson (c.f.~Fig.~\ref{fig:neutralinodecay})
are dominant \cite{Mukhopadhyaya:1998xj}. The partial decay widths read
\begin{subequations}
\begin{align}
   \Gamma \left( \chi_1^0 \rightarrow W^\pm l^\mp \right)
=& \frac{G_F m_{\chi_1^0}^3}{4 \sqrt 2 \pi} \sum_i
   \left| V_{1i \ \text{LO}}^{(\chi,e)} \right|^2 f_W (m_{\chi_1^0} )
   \left(1+\mathcal O \left(s_{2\beta}\frac{m_Z^2}{\mu^2}\right)\right) \ ,
\\ \Gamma\left( \chi_1^0 \rightarrow Z \nu \right)
=& \frac{G_F m_{\chi_1^0}^3}{2 \sqrt 2 \pi} \sum_i
   \left| V_{1i\ \text{LO}}^{(\chi,\nu)} \right|^2 f_Z (m_{\chi_1^0})
   \left( 1 + \mathcal O \left( s_{2\beta}\frac{m_Z^2}{\mu^2} \right) \right)
\ ;
\end{align}
\label{Neutralino_Decay_Width}
\end{subequations}
here $V^{(\chi,e)}_{1i\ \text{LO}}$ and $V^{(\chi,\nu)}_{1i\ \text{LO}}$
are the charged and neutral current matrix elements at leading order,
which are given in Eqs.~(\ref{meneutral}) and (\ref{mecharged}),
respectively.
The function $f_{W,Z}$ is a phase space factor which becomes important for
neutralino masses close to the lower bound of \unit[100]{GeV}
(cf.~Fig.~\ref{fig:phasespace}),
\begin{align}
f_{W,Z}(m_{\chi^0_1})
  = \left(1 - \frac{m_{W,Z}^2}{m_{\chi^0_1}^2}\right)^2
    \left(1 + 2 \frac{m_{W,Z}^2}{m_{\chi^0_1}^2}\right) \ .
\end{align}
For large NLSP masses, $m_{\chi^0_1} \gg m_Z$, one has
\begin{align}
   BR \left(\chi_1^0 \rightarrow W^{\pm}l^{\mp}\right)
 \simeq
   2\ BR\left(\chi_1^0 \rightarrow Z\nu\right) \ ,
\end{align}
whereas in the region $m_{\chi^0_1} \simeq \unit[100]{GeV}$
\begin{align}
  BR\left(\chi_1^0 \rightarrow W^{\pm}l^{\mp}\right)
 \simeq
  5\ BR\left(\chi_1^0 \rightarrow Z\nu\right)
\ .
\end{align}

The total neutralino NLSP decay width is given by the sum
\begin{align}
   \Gamma_{\chi_1^0}
 = \Gamma(\chi_1^0 \rightarrow W^\pm l^\mp)
 + \Gamma(\chi_1^0 \rightarrow Z \nu)
\ ,
\end{align}
which corresponds to the decay length \cite{Bobrovskyi:2010ps}
\begin{align}
c \tau_{\chi_1^0} \gtrsim \ \unit[2.7]{m}
&\left(\frac{m_{\chi_1^0}}{\unit[100]{GeV}} \right)^{-1}
\left(\frac{\zeta}{3\times 10^{-8}}\right)^{-2} \nonumber \\
&\times\left(2 f_W(m_{\chi_1^0})+f_Z(m_{\chi_1^0})\right)^{-1}
\left( 1+\mathcal O \left(s_{2\beta}\frac{ m_Z^2}{ \mu^2} \right) \right) \ .
\label{minneutralino}
\end{align}
Note that both, the gravitino and the neutralino NLSP lifetimes are functions
just of $\zeta$ and the masses, without any further parameters.
This is the case for the standard supergravity mass spectrum with
a bino-like NLSP. This direct connection between the gravitino and NLSP
lifetimes is the basis of our analysis.

\section{Decaying neutralino NLSP at the LHC}
\label{sec:signatures_and_reach}

In this section we classify the main LHC signatures of decaying neutralinos.
We show that for small values of the R-parity breaking parameter~$\zeta$ usual SUSY searches are insufficient to find the signal.
However, R-parity violation leads to new signals including striking secondary vertices at large distances from the primary interaction point.

\subsection{Decay signatures}
\label{sec:signatures}

\begin{figure}
\centering
\psfrag{Wp}{\small$W^+$}
\psfrag{e}{\small$e^-$}
\psfrag{q}{\small$q$}
\psfrag{sq}{\small$\widetilde q$}
\psfrag{gl}{\small$\widetilde g$}
\psfrag{Z}{\small$Z$}
\psfrag{nu}{\small$\nu$}
\psfrag{mm}{\small$\mu^-$}
\psfrag{mp}{\small$\mu^+$}
\psfrag{ch}{\small$\chi^0_1$}
\includegraphics[width=0.7\textwidth]{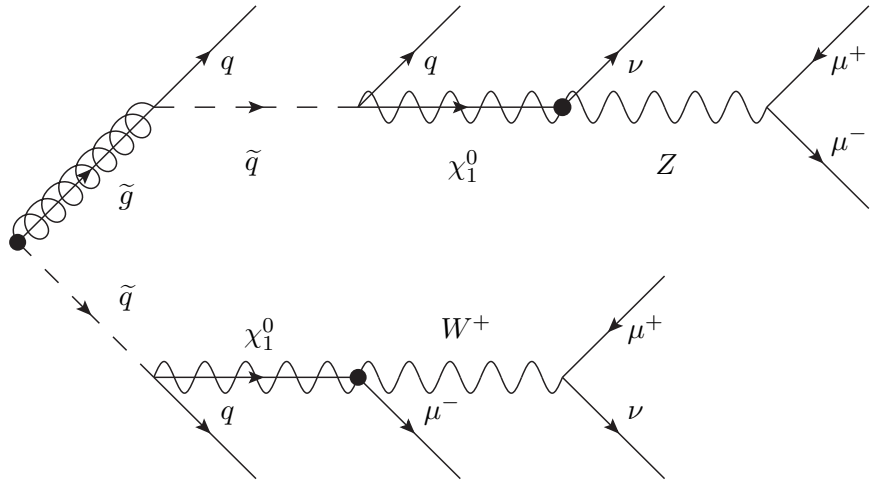}
\caption{Typical R-parity violating decay chain at the LHC. The primary vertex and the secondary vertices are highlighted.}
\label{fig:decaychain}
\end{figure}

Consider for simplicity, the following cascade process:
\begin{equation}
   qg \rightarrow \widetilde q \widetilde g
 \rightarrow
   j j j \chi^0_1 \chi^0_1
\;,
\end{equation}
where $\widetilde q$ is a squark, $\widetilde g$ is a gluino, and $j$ denotes a jet.
The final state neutralinos decay in a secondary vertex into  $W$-bosons and leptons as well as into $Z$-bosons and neutrinos.
Fig.~\ref{fig:decaychain} shows an example of a decay cascade with muons in final state.
The distance between the collision point and the secondary vertex depends on the decay width of the neutralino~\eqref{Neutralino_Decay_Width} and hence on the R-parity breaking parameter~$\zeta$.

Table~\ref{tab:finstates} summarizes the LHC signatures for sufficiently large values of $\zeta$ such that it is probable that both neutralinos decay inside of the tracker volume.
All the signatures contain at least three jets from the antecedent supersymmetric decays.
The signatures are classified according to the final states in the neutralino decays: leptonic signatures involving only leptons in the final state, semi-leptonic signatures involving at least two charged leptons and jets, single lepton signatures containing only one lepton, all-hadronic signatures where only jets accompanied by neutrinos are present, and finally invisible channels where both neutralinos decay solely to neutrinos.
Additionally we single out channels having a considerable amount of missing transverse energy $\slashed E_T$ from $Z$-boson decays, since $\slashed E_T$ is one of the main features searched for in usual searches for new physics.
Furthermore the channels labeled as \emph{opposite sign} could be found in usual supersymmetry (SUSY) searches as they include a considerable amount of $\slashed E_T$, many jets and one isolated lepton pair with different signs.
However, some searches remove events with muon pairs having invariant mass around the $Z$ pole in order to dispose of Drell-Yan $Z/\gamma^*\rightarrow l \bar l$ processes.
Note that in the model presented in this work this cut would lead to a suppression of the signal.
Other leptonic and semi-leptonic channels also contain opposite-sign lepton pairs but only small amount of $\slashed E_T$ and therefore they are not considered in the usual searches, (c.f.~\cite{Aad:2011xm,Chatrchyan:2011bz}).
Neutralino decays lead also to signatures containing same-sign lepton pairs but since no $\slashed E_T$ is present in these channels they are usually discarded in order to suppress various backgrounds \cite{Chatrchyan:2011wb}.

\begin{table}
\centering
\begin{tabular}{l r@{ $\rightarrow$ }l r}
   \toprule
   category
 & \multicolumn{2}{ c }{$\chi^0_1$ decays}
 & LHC signature
\\ \midrule
   leptonic
 & $ W^+ W^-  l \bar l $
 & $ \bar l l l \bar l \nu \nu $
 & \multirow{6}{*}{$ 3 j + 2 l + 2 \bar l + \slashed E_T $}
\\
 & $ W^+ W^+  ll $
 & $ \bar l \bar l l l  \nu \nu $
 &
\\
 & $ W^- W^-  \bar l \bar l $
 & $ l l  \bar l \bar l \nu \nu $
 &
\\
 & $ Z W^-  \bar l \nu $
 & $ l \bar l l \bar l \nu \nu $
 &
\\
 & $ Z W^+ l \nu $
 & $ l \bar l \bar l l \nu \nu $
 &
\\
 & $ Z Z \nu \nu $
 & $ l \bar l l \bar l \nu \nu $
 &
\\ \cmidrule{2-4}
   \emph{(opposite sign,}
 & $ Z W^+ l \nu $
 & $ \nu \nu  \bar l l \nu \nu $
 &
\\ \emph{$\slashed E_T $ from $Z$)}
 & $ Z W^- \bar l \nu $
 & $ \nu \nu  l \bar l \nu \nu $
 & $ 3 j + 1 l + 1 \bar l + \slashed E_T $
\\
 & $ Z Z \nu \nu $
 & $ \nu \nu  l \bar l \nu \nu $
 &
\\ \midrule
   semi-leptonic
 & $ W^+ W^-  l \bar l $
 & $ j j l l \bar l  \nu $
 &
\\
 & $ W^+ W^+  l l $
 & $ j j \bar l  l l  \nu $
 & $ 5 j + 2 l + 1 \bar l + \slashed E_T $
\\
 & $ Z W^+  l \nu $
 & $ l \bar l j j l \nu $
 &
\\ \cmidrule{2-4}
 & $ W^+ W^-  l \bar l $
 & $ j j \bar l l \bar l  \nu $
 &
\\
 & $ W^- W^-  \bar l \bar l $
 & $ j j l  \bar l \bar l  \nu $
 & $ 5 j + 1 l + 2 \bar l + \slashed E_T $
\\
 & $ Z W^-  \bar l \nu $
 & $  l  \bar l jj  \bar l  \nu $
 &
\\ \cmidrule{2-4}
 & $ Z W^+  l \nu $
 & $ j j \bar l l  \nu \nu $
 &
\\
 & $ Z W^-  \bar l \nu $
 & $ jj l \bar l \nu  \nu $
 & $ 5 j + 1 l + 1\bar l + \slashed E_T $
\\
 & $ Z Z \nu \nu $
 & $ jj l \bar l \nu  \nu $
 &
\\ \cmidrule{2-4}
 & $ W^+ W^- l \bar l $
 & $ j j j j l  \bar l $
 & $7 j + 1l + 1\bar l \phantom{ \; + \slashed E_T } $
\\ \cmidrule{2-4}
  \emph{(same sign,}
 & $ W^+ W^+  l l $
 & $ j j j j l l $
 & $7 j + 2 l \phantom{ \; + \slashed E_T } $
\\ \emph{no $\slashed E_T $)}
 & $ W^- W^-  \bar l \bar l $
 & $ j j j j \bar l \bar l $
 & $7 j + 2 \bar l \phantom{ \; + \slashed E_T } $
\\ \midrule
   single lepton
 & $ Z W^+  l \nu $
 & $ j j j j l \nu$
 & $ 7 j + 1 l  + \slashed E_T $
\\ \emph{($\slashed E_T$ from Z)}
 & $ Z W^-  \bar l \nu $
 & $ j j j j \bar l \nu$
 & $ 7 j + 1 \bar l + \slashed E_T $
\\ \cmidrule{2-4}
 & $ Z W^+  l \nu $
 & $ \nu \nu j j l \nu$
 & $ 5 j + 1 l + \slashed E_T $
\\
 & $ Z W^-  \bar l \nu $
 & $ \nu \nu j j \bar l \nu$
 & $ 5 j + 1 \bar l + \slashed E_T $
\\ \midrule
   all-hadronic
 & $ Z Z \nu \nu $
 & $ j j j j \nu \nu $
 & $ 7 j + \slashed E_T $
\\ \emph{($ \slashed E_T $ from Z)}
 & $ Z Z \nu \nu $
 & $ \nu \nu j j \nu \nu $
 & $ 5 j + \slashed E_T $
\\ \midrule
   invisible
 & $ Z Z \nu \nu $
 & $ \nu \nu \nu \nu \nu \nu $
 & $ 3 j + \slashed E_T $
\\ \emph{($ \slashed E_T $ from 2 Z)}
 & \multicolumn{2}{c}{}
 &
\\ \bottomrule
\end{tabular}
\caption{Possible final states if both neutralinos decay inside the tracking volume.}
\label{tab:finstates}
\end{table}
\begin{table}
\centering
\begin{tabular}{l r@{ $\rightarrow$ }l r}
  \toprule
   category
 & \multicolumn{2}{c}{$\chi^0_1$ decays}
 & LHC signature
\\ \midrule
   leptonic
 & $W^+ l $\
 & $\bar l l \nu $
 &
\\ \emph{(opposite sign)}
 & $W^- \bar l $
 & $l \bar l \nu $
 & $3 j + 1 l + 1 \bar l + \slashed E_T $
\\
 & $Z \nu $
 & $l \bar l \nu$
 &
\\ \midrule
   single lepton
 & $W^+ l $
 & $j j l $
 & $5 j + 1 l + \slashed E_T $
\\
 & $ W^- \bar l $
 & $ jj \bar l $
 & $ 5 j + 1 \bar l + \slashed E_T $
\\ \midrule
   all-hadronic
 & $Z \nu $
 & $j j \nu $
 & $5 j + \slashed E_T $
 \\ \midrule
   invisible
 & $Z \nu $
 & $\nu \nu \nu$
 & $3 j + \slashed E_T $
\\ \bottomrule
\end{tabular}
\caption{Possible final states if one of the neutralinos decays outside the tracking volume.}
\label{tab:finstatesout}
\end{table}

If the value of $\zeta$ is rather small one of the neutralinos will decay outside of the detector leading to  signatures with large amount of $\slashed E_T$ as shown in Table~\ref{tab:finstatesout}.
The leptonic decays of one of the neutralinos inside the detector lead to a perfect opposite-sign signature.
As mentioned above this signature can be hidden if one rejects events where the invariant mass distribution of the lepton pair is in the range of the $Z$-boson mass.
Another strategy is the search for single lepton events with large amount of missing transverse energy.

Thus the applicability and the reach of the usual SUSY searches applied to the model presented in this work depends crucially on the size of R-parity breaking.
In order to further evaluate this statement we investigated a number of characteristic variables in supersymmetric events.
The events were generated with \software{pythia} as described in the next section,
with the mSUGRA boundary conditions $m_{\nicefrac{1}{2}} = m_{0}= 270$, $\tan \beta = 10$, $a_0 = 0$, and $\mu>0$.
R-parity violating neutralino decays were taken into account.

\begin{figure}
\centering
\psfrag{Events}{\small\hspace{-6.5em} Neutralinos/$\unit[10]{fb^{-1}}$}
\psfrag{gb}{\small$\beta\gamma$}
\psfrag{0}{\footnotesize\hspace{-0.1em}0}
\psfrag{1}{\footnotesize\hspace{-0.1em}1}
\psfrag{2}{\footnotesize\hspace{-0.1em}2}
\psfrag{3}{\footnotesize\hspace{-0.1em}3}
\psfrag{4}{\footnotesize\hspace{-0.1em}4}
\psfrag{5}{\footnotesize\hspace{-0.1em}5}
\psfrag{6}{\footnotesize\hspace{-0.1em}6}
\psfrag{7}{\footnotesize\hspace{-0.1em}7}
\psfrag{8}{\footnotesize\hspace{-0.1em}8}
\psfrag{9}{\footnotesize\hspace{-0.1em}9}
\psfrag{10}{\footnotesize\hspace{-0.1em}10}
\psfrag{o}{\footnotesize\hspace{-0.4em}0}
\psfrag{200}{\footnotesize\hspace{-0.4em}200}
\psfrag{400}{\footnotesize\hspace{-0.4em}400}
\psfrag{600}{\footnotesize\hspace{-0.4em}600}
\psfrag{800}{\footnotesize\hspace{-0.4em}800}
\psfrag{1000}{\footnotesize\hspace{-0.4em}1000}
\psfrag{1200}{\footnotesize\hspace{-0.4em}1200}
\includegraphics[width=0.7\textwidth]{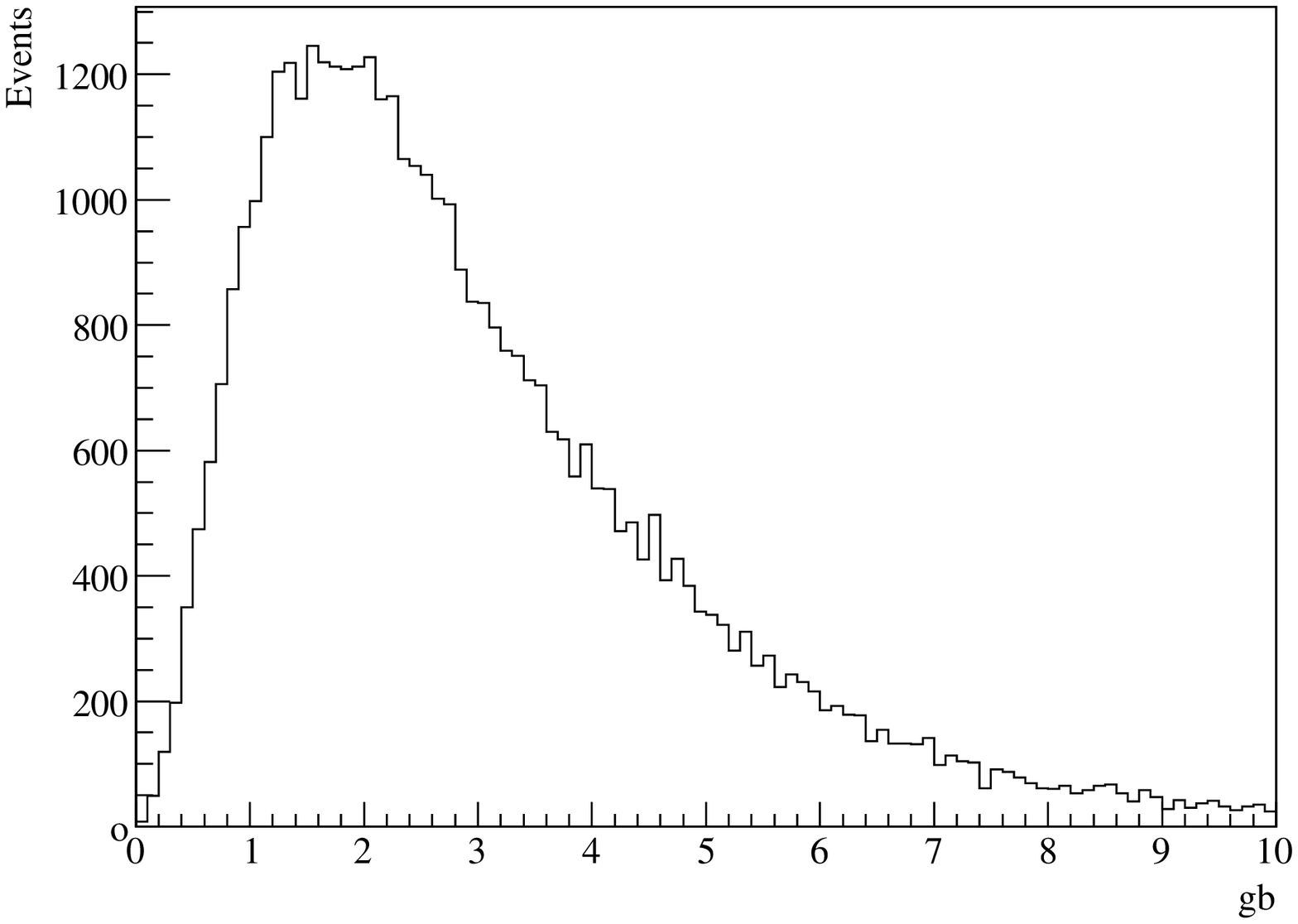}
\caption{$\beta \gamma$ distribution of neutralinos at generator level
for benchmark point HH27 (see Table~\ref{tab:parameter_point_definition}).
The number of neutralinos corresponds to twice the number of the events scaled to $\unit[10]{fb^{-1}}$ at $\sqrt{s} = \unit[7]{TeV}$.}
\label{fig:gammabeta_distribution}
\end{figure}
\begin{figure}
\centering
\psfrag{Events}{\small\hspace{-4.5em} Events/$\unit[10]{fb^{-1}}$}
\psfrag{MPT \(GeV\)}{\small\hspace{3.5em}$\slashed p_T$}
\psfrag{0}{\footnotesize0}
\psfrag{o}{\footnotesize\hspace{-0.05em}0}
\psfrag{200}{\footnotesize\hspace{-0.05em}200}
\psfrag{400}{\footnotesize\hspace{-0.05em}400}
\psfrag{600}{\footnotesize\hspace{-0.05em}600}
\psfrag{800}{\footnotesize\hspace{-0.05em}800}
\psfrag{1000}{\footnotesize\hspace{-0.05em}1000}
\psfrag{1200}{\footnotesize\hspace{-0.05em}1200}
\psfrag{500}{\footnotesize\hspace{-0.2em}500}
\psfrag{100o}{\footnotesize\hspace{-0.2em}1000}
\psfrag{1500}{\footnotesize\hspace{-0.2em}1500}
\psfrag{2000}{\footnotesize\hspace{-0.2em}2000}
\psfrag{2500}{\footnotesize\hspace{-0.2em}2500}
\psfrag{3000}{\footnotesize\hspace{-0.2em}3000}
\psfrag{3500}{\footnotesize\hspace{-0.2em}3500}
\psfrag{Z = 3e-8}{\footnotesize$\zeta = 3\times10^{-8}$}
\psfrag{Z = 1e-9}{\footnotesize$\zeta = 1\times10^{-9}$}
\psfrag{Z = 0}{\footnotesize$\zeta = 0$}
\includegraphics[width=0.7\textwidth]{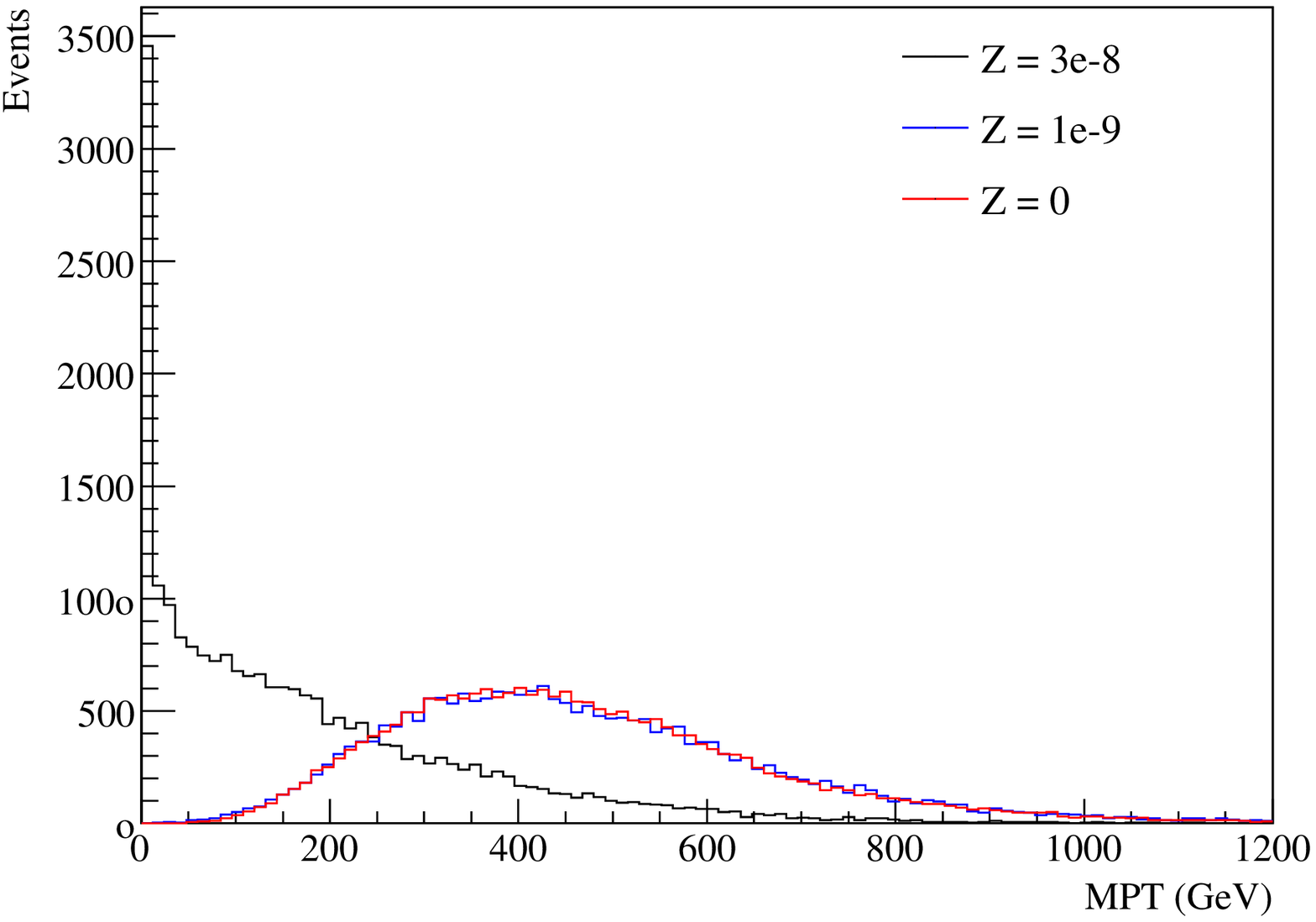}
\caption{$\slashed p_T$ distribution at generator level for benchmark point
HH27 (see Table~\ref{tab:parameter_point_definition}) and different values of
the R-parity breaking parameter $\zeta$.
Generator level $\slashed p_T$ is defined as sum over the $p_T$ of i) neutralinos decaying outside of the detector (see Section~\ref{sec:physreco}) and ii) all neutrinos produced inside of the detector.
The number of events is scaled to $\unit[10]{fb^{-1}}$ at $\sqrt{s} = \unit[7]{TeV}$.
}
\label{fig:MET}
\end{figure}

Fig.~\ref{fig:gammabeta_distribution} shows the distribution of the $\beta \gamma$ factors of the neutralinos.
This factor enters the formula for the neutralino decay length and one sees from the plot that analytic results in the literature, which have been computed
with $\beta \gamma = 1$, are correct within one order of magnitude.
The most important kinematic property connected with the neutralino decay length is the amount of missing transverse momentum $\slashed p_T$ which is shown in Fig.~\ref{fig:MET} for different values of the R-parity violation parameter $\zeta$.
The missing transverse momentum was computed as the sum of the transverse momenta of all neutrinos produced in the detector before the hadronic calorimeter ($r < \unit[1800]{mm}$, $|z| < \unit[3700]{mm}$) and the transverse momenta of the neutralinos decaying outside the hadronic calorimeter.
The $\slashed p_T$ distribution of the R-parity conserving model $\zeta = 0$ cannot be distinguished from the model with $\zeta = 1\times 10^{-9}$.
However, the distribution is significantly different for $\zeta = 3\times 10^{-8}$ since in this case most events have only very little missing transverse momentum due to early neutralino decays.
This suggests that our model could only hardly be discovered in usual searches relying on $\slashed E_T$.
A further analysis with full detector simulation is needed in order to properly evaluate the discovery potential of usual SUSY searches.

\begin{figure}
%\vspace{-5ex} % there seems to be an error in the bounding box
\centering
\psfrag{Nm}{\footnotesize\hspace{-2em}N ($\mu^{\pm}$)}
\psfrag{Ne}{\footnotesize\hspace{-2em}N ($e^{\pm}$)}
\psfrag{0}{\scriptsize\hspace{-0.05em}0}
\psfrag{2}{\scriptsize\hspace{-0.05em}2}
\psfrag{4}{\scriptsize\hspace{-0.05em}4}
\psfrag{6}{\scriptsize\hspace{-0.05em}6}
\psfrag{8}{\scriptsize\hspace{-0.05em}8}
\psfrag{1}{\scriptsize\hspace{-0.1em}1}
\psfrag{2}{\scriptsize\hspace{-0.05em}2}
\psfrag{3}{\scriptsize\hspace{-0.05em}3}
\psfrag{5}{\scriptsize\hspace{-0.05em}5}
\psfrag{7}{\scriptsize\hspace{-0.05em}7}
\psfrag{9}{\scriptsize\hspace{-0.05em}9}
\psfrag{10}{\scriptsize\hspace{-0.2em}10}
\psfrag{100}{\scriptsize\hspace{-0.3em}$10^2$}
\psfrag{1000}{\scriptsize\hspace{0.1em}$10^3$}
\psfrag{10000}{\scriptsize\hspace{0.2em} $10^4$}
\psfrag{Events}{\footnotesize\hspace{-5.5em} Events/$\unit[10]{fb^{-1}}$}
\psfrag{Z = 3e-8}{\scriptsize$\zeta = 3\times10^{-8}$}
\psfrag{Z = 0}{\scriptsize$\zeta = 0$}
\subfloat[Number of electrons.]{\label{fig:number_of_electrons}\includegraphics[width=0.49\textwidth]{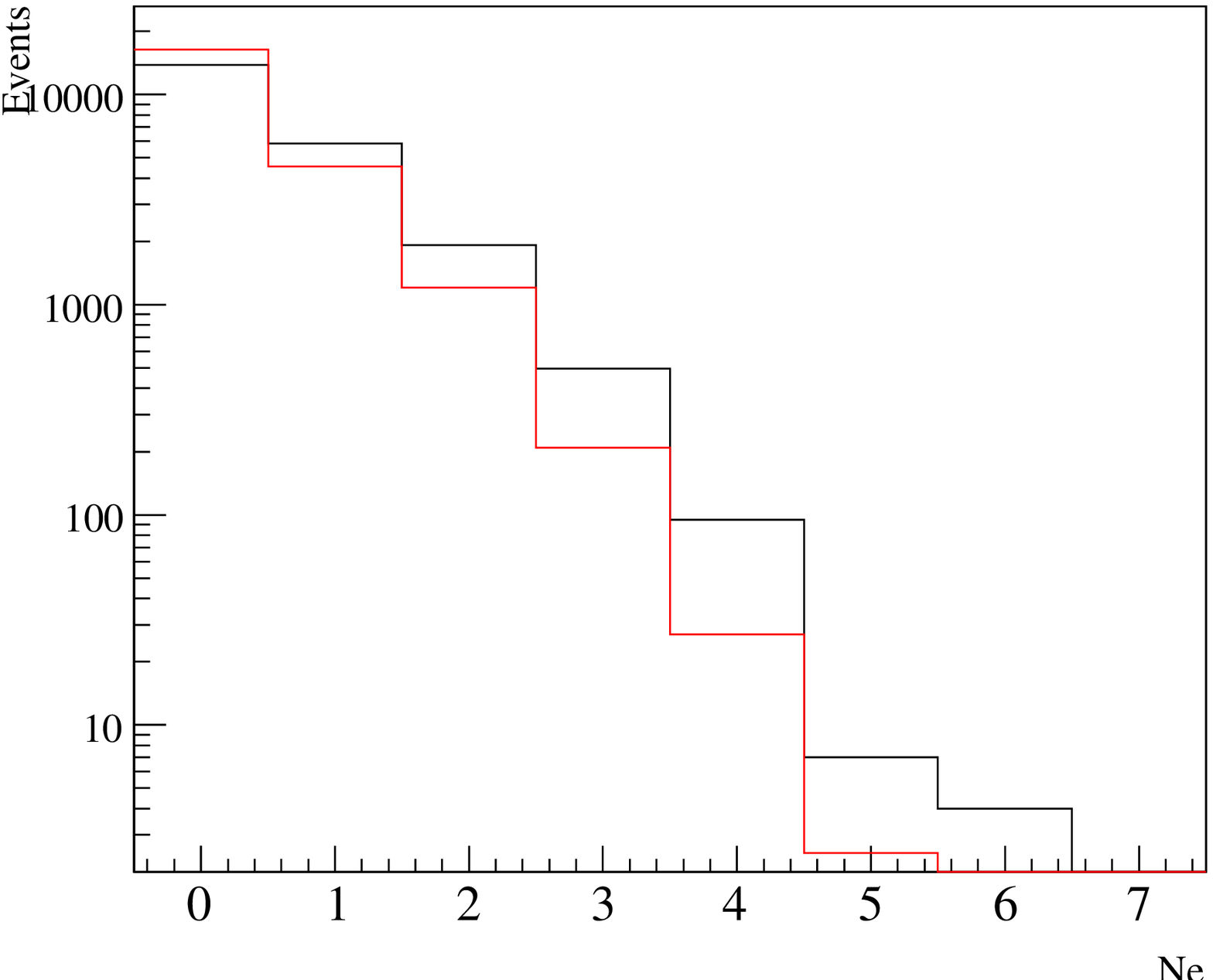}}
\subfloat[Number of muons.]{\label{fig:number_of_muons}\includegraphics[width=0.49\textwidth]{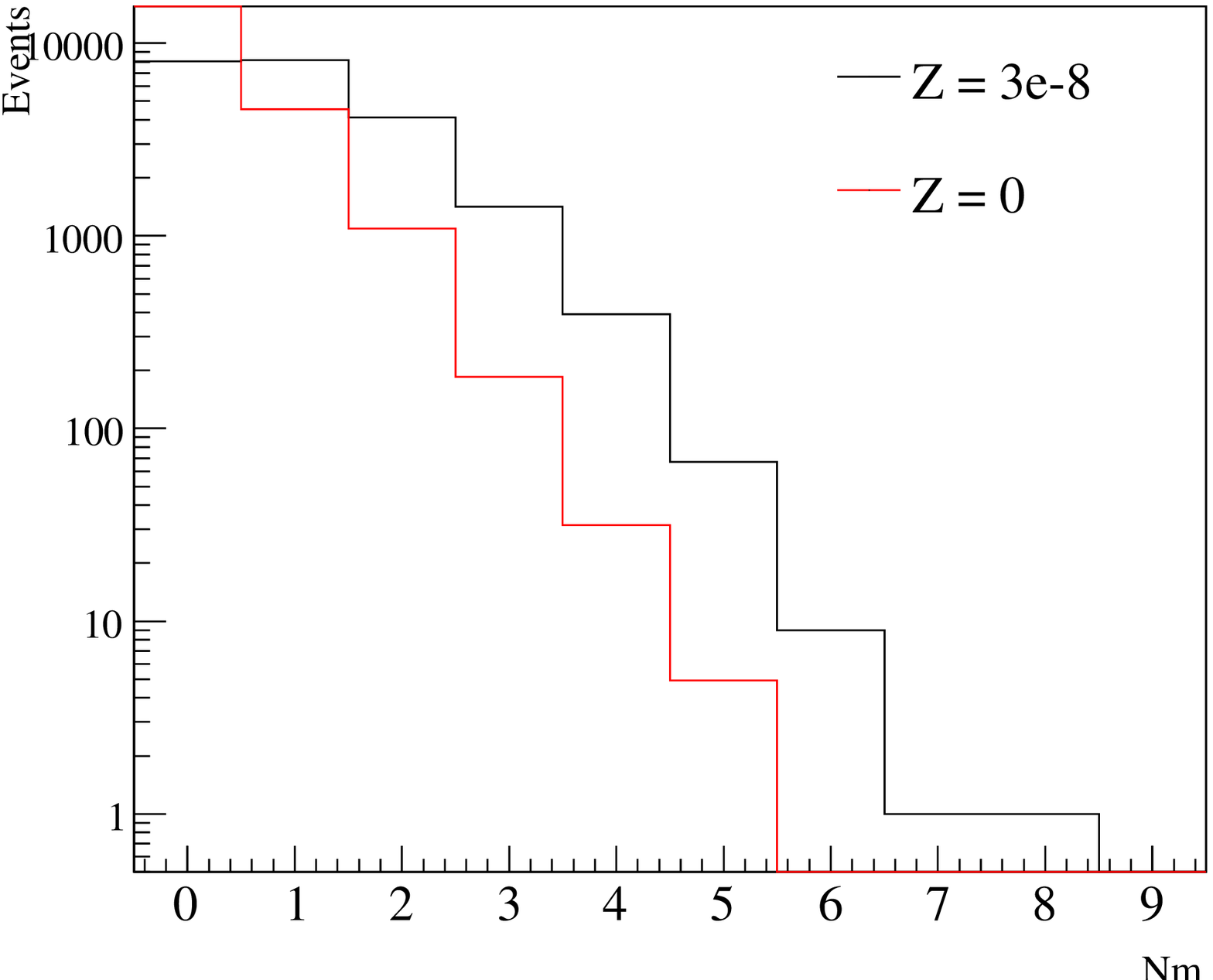}}
\caption{The number of generated particles per event after selection cuts described in Table~\ref{tab:gencuts}.
The color code for the curves in both plots is given in Fig.~\ref{fig:number_of_muons}.
The number of events is scaled to $\unit[10]{fb^{-1}}$ at $\sqrt{s} = \unit[7]{TeV}$.
}
\label{fig:number_of_leptons}
\end{figure}

Another general feature of models with relatively large $\zeta$ is the large possible number of leptons in the final state, illustrated in Fig.~\ref{fig:number_of_leptons}.
The generator level particles selected for this plot had to fulfill the criteria shown in Table~\ref{tab:gencuts} imposed in order to select leptons from hard processes which could be reconstructed in a realistic detector.
The cuts on the vertex position represent a pessimistic estimate of the reconstruction efficiency (see~Section~\ref{sec:mueff}).

\begin{table}
\centering
\begin{tabular}{lccrl}
  \toprule
    particle
  & transverse momentum
  & pseudorapidity
  & \multicolumn{2}{c}{vertex position}
 \\ \midrule
    electron
  & $p_T > \unit[7]{GeV}$
  & $|\eta| < 2.5$
  & $r < \unit[400]{mm}$
  & $|z| < \unit[1300]{mm}$
 \\ muon
  & $p_T > \unit[6]{GeV}$
  & $|\eta| < 2.5$
  & $r < \unit[4000]{mm}$
  & $|z| < \unit[6000]{mm}$
 \\ \bottomrule
\end{tabular}
\caption{Cuts for the generator level particle selection for the study of particle multiplicity.}
\label{tab:gencuts}
\end{table}

\subsection{Search strategies}
\label{sec:search_strategies}

As mentioned in the previous section one of the striking features of the presented model are events with secondary vertices and possibly many leptons in the final state.
The search for a secondary vertex is crucial in order to ensure the R-parity violating nature of the decays.
Possible search strategies can be optimized in order to find some of the channels described in Tables~\ref{tab:finstates} and~\ref{tab:finstatesout}.
It is remarkable that many channels allow for the full reconstruction of the neutralino mass: all decay chains including $Z$-bosons or hadronically decaying $W$-bosons.
The reconstruction of the neutralino mass from the particles produced in the $Z$-boson decay depends crucially on the full reconstruction of the secondary vertex, which is beyond the scope of this work\footnote{The four-vector pointing to the secondary vertex and the three-momenta of the leptons or jets from the $Z$-boson provide sufficient information for the reconstruction of the neutralino mass.}.
This method of neutralino mass reconstruction works also in R-parity conserving models where the neutralino decays into $Z$-boson and gravitino \cite{Meade:2010ji}.

For example, one promising search strategy working for all $\zeta$ values considered in this work is based on \emph{single lepton} events with some number of hard jets and missing transverse energy larger than \unit[90]{GeV}.
After the preselection one could look for events where the lepton is coming from a secondary vertex and try to reconstruct the $W$-boson mass from a jet pair.
In the final step one could try to reconstruct the neutralino mass from the jets selected in the previous step and the lepton.
However such study depends crucially on the knowledge of the detector response in the case of late decaying particles.
A neutralino can decay in various detector components and lead to unusual signals.
Furthermore the mass resolution is limited by the uncertainty in the jet energy scale and by the uncertainty in the determination of the jet momentum direction.

We will focus our study on leptonic final states, which have a particularly
clean signature, and reconstruct the $Z$-boson coming from a secondary vertex.
We will use only muon and track objects for which we assume to have modeled a realistic detector response (see Section~\ref{sec:mueff}).
A possible background for this search are cosmic muons leaving no track in the detector.
It is important to note that one would miss the signal in this channel entirely if one imposes a cosmic muon veto which rejects all events with muon pairs having no associated tracks (c.f.~\cite{Aad:2011xm}).

\section{Simulation of signal and background}
\label{sec:signal_and_background}

In this section we define a set of representative points in the parameter space of our model and describe the generation of the signal and dominant standard model (SM) background samples.
In particular we examine the simulation of detector effects using the generic detector simulation \software{delphes} 1.8 \cite{Ovyn:2009tx} on signal and background in the presence of secondary vertices.

\subsection{Benchmark points}
\label{sec:Points}

A typical set of boundary conditions for the supersymmetry breaking parameters of the MSSM at the grand unification scale is given by equal scalar and gaugino masses, $m_0 = m_{\nicefrac{1}{2}}$.
These boundary conditions lead to a bino-like neutralino $\chi^0_1$ as NLSP.
We choose a representative value of $\tan \beta$ and set the scalar trilinear couplings to zero,
\begin{equation}
   a_0
 = 0
\;,
\quad
   \tan \beta
 = 10
\;.
\end{equation}
Thus the universal gaugino mass remains the only independent supersymmetry breaking parameter which will be varied in the present study.

Electroweak precision tests lead to lower bounds on the supersymmetric particle spectrum.
The LEP lower bound for the lightest Higgs at $\tan \beta = 10$ is $m_H \gtrsim \unit[95]{GeV}$ \cite{Nakamura:2010zzi}.
In the present study the lightest superparticle spectrum corresponds to the choice $m_0 = m_{\nicefrac{1}{2}} =\unit[270]{GeV}$ (HH27).
At this benchmark point the NLSP is a neutralino with mass $m_{\chi_1^0} = \unit[105.8]{GeV}$ and the lightest Higgs boson has a mass $m_h =  \unit[110.4]{GeV}$.
\begin{table}
\centering
\begin{tabular}{lrccccc}
   \toprule
 & \multicolumn{2}{c}{GUT masses}
 & \multicolumn{4}{c}{particle masses}
\\ \cmidrule(r){2-3}\cmidrule(l){4-7}
 & $m_0$
 & $m_{\nicefrac{1}{2}}$
 & $m_{\chi_0^1}$
 & $m_h$
 & \multicolumn{1}{c}{$m_{\widetilde g}$}
 & \multicolumn{1}{c}{$m_{\widetilde u}$}
\\ \midrule
   HH27
 & 270
 & 270
 & 105.8
 & 110.5
 & 662.4
 & 653.4
\\ HH35
 & 350
 & 350
 & 140.5
 & 112.5
 & 841.7
 & 831.8
\\ HH50
 & 500
 & 500
 & 205.7
 & 115.1
 & 1170
 & 1160
\\ HH65
 & 650
 & 650
 & 271.5
 & 116.7
 & 1492
 & 1481
\\ HH80
 & 800
 & 800
 & 337.8
 & 117.9
 & 1809
 & 1798
\\ \bottomrule
\end{tabular}
\caption{Definition of the benchmark points together with some particle masses; all masses are in $\unit{GeV}$.}
\label{tab:parameter_point_definition}
\end{table}
\begin{table}
\centering
\begin{tabular}{cccccc}
   \toprule
 & \multicolumn{4}{c}{partial crosssections [\unit{fb}]}
 &
\\ \cmidrule(lr){2-5}
 & $\sigma(\widetilde q \widetilde g)$
 & $\sigma(\widetilde q \widetilde q)$
 & $\sigma(\widetilde q \overline{\widetilde q})$
 & $\sigma(\widetilde g \widetilde g)$
 & $\sigma(\text{tot})$ [\unit{fb}]
\\ \midrule
   \multirow{2}{*}{HH27}
 & 1090
 & 682
 & 256
 & 208
 & 2236
\\

 & (739)
 & (570)
 & (174)
 & (83)
 & (1566) \bigstrut[b]
\\ \multirow{2}{*}{HH35}
 & 172 \bigstrut[t]
 & 149
 & 38
 & 26
 & 385
\\
 & (105)
 & (126)
 & (25.2)
 & (8.47)
 & (265) \bigstrut[b]
\\ \multirow{2}{*}{HH50}
 & 8.91 \bigstrut[t]
 & 11.8
 & 1.7
 & 0.95
 & 23.36
\\
 & (4.36)
 & (10.1)
 & (1.02)
 & (0.206)
 & (15.7) \bigstrut[b]
\\ \multirow{2}{*}{HH65}
 & 0.579 \bigstrut[t]
 & 1.01
 & 0.0943
 & 0.0466
 & 1.73
\\
 & (0.216)
 & (0.877)
 & (0.0458)
 & ($6.37 \times 10^{-3}$)
 & (1.145) \bigstrut[b]
\\ \multirow{2}{*}{HH80}
 & 0.0379 \bigstrut[t]
 & 0.0805
 & $5.37 \times 10^{-3}$
 & $2.44 \times 10^{-3}$
 & 0.126
\\
 & (0.0109)
 & (0.0723)
 & ($1.98 \times 10^{-3}$)
 & ($0.203 \times 10^{-3}$)
 & (0.0854)
\\ \bottomrule
\end{tabular}
\caption{Production cross sections at NLO (LO) at the benchmark points calculated with \software{prospino}.}
\label{tab:crosssections}
\end{table}

In order to probe the region of gluino and squark masses accessible at the LHC~\cite{Ball:2007zza} we increase the gaugino mass parameter in four steps: $m_{\nicefrac{1}{2}} = 350$, $500$, $650$, $\unit[800]{GeV}$.
Some particle masses at these points are shown in Table~\ref{tab:parameter_point_definition}. For the different benchmark points the production cross sections, calculated with \software{prospino} 2.1 \cite{Beenakker:1996ed} at $\sqrt{s} = \unit[7]{TeV}$, are given in
Table~\ref{tab:crosssections}.

For the R-parity breaking parameter~$\zeta$ we choose the following values: $\zeta = 3 \times 10^{-8}$~\cite{Bobrovskyi:2010ps}, $\zeta = 2\times 10^{-8}$~\cite{Vertongen:2011mu}, $1\times 10^{-8}$, $5\times 10^{-9}$, $1\times 10^{-9}$, $5\times 10^{-10}$ and $1\times 10^{-10}$. Note that for gravitino masses $m_{\nicefrac{3}{2}} = \mathcal{O}(\unit[100]{GeV)}$ one has $\zeta \lesssim 1\times 10^{-9}$~\cite{Bobrovskyi:2010ps}.

\subsection{Major backgrounds}

Neutralino decays always have $W$- and $Z$-bosons in the final state (c.f.~Fig.~\ref{fig:neutralinodecay}). In our study we focus on the reconstruction of $Z$-boson decays to muon pairs. Therefore we only consider SM backgrounds which lead to at least two muons in the final state originating from $W$- or $Z$-bosons:
\begin{itemize}
 \item $t\bar t$ production: $W$-bosons from top quark decays.
 \item $Z$ production
 \item Di-boson production ($WW$, $WZ$, $ZZ$)
 \item Tri-boson production ($WWW$, $WWZ$, $ZZW$, $ZZZ$)
\end{itemize}
Table~\ref{tab:backgrounds} gives an overview of the background samples used in our analysis.
We assume that pure QCD background can be efficiently suppressed in multi-lepton final states with high transverse momentum, particularly after imposing lepton isolation criteria (c.f.~\cite{Aad:2009wy,Chatrchyan:2011bz}).

\begin{table}
\centering
\subfloat[Samples of SM background.
The production cross sections are taken from \cite{Kidonakis:2010bb, Binoth:2008kt, Campbell:2011bn, Gavin:2010az}.]
{
\begin{tabular}[t]{lrr}
   \toprule Sample
 & $\sigma$ [\unit{pb}]
 & events
\\ \midrule
 $t\bar{t}$
 & 163
 & 200000 \bigstrut[b]
\\
 $Z$ \bigstrut[t]
 & 977
 & 700000 \bigstrut[b]
\\
 $W^+W^-$ \bigstrut[t]
 & 47
 & 50000
\\
 $ZZ$
 & 6.46
 & 50000
\\
  $W^+Z$
 & 11.88
 & 50000 \bigstrut[b]
\\
  $W^-Z$
 & 6.69
 & 50000 \bigstrut[b]
\\
 $W^+W^-Z$ \bigstrut[t]
 & 0.182
 & 15000
\\
 $W^+ZZ$
 & 0.040
 & 15000
\\
 $W^+W^-W^+$
 & 0.146
 & 15000
\\
 $ZZZ$
 & 0.015
 & 5629
\\ \bottomrule
\end{tabular}
\label{tab:backgrounds}
}
\hspace{0.1\textwidth}
\subfloat[Samples of signal events for different benchmark points (see Table~\ref{tab:parameter_point_definition}) and $\zeta = \alpha \times 10^{-9}$ ($\alpha=0.1$, $0.5$, $1,5$, $10$, $20$, $30$).]
{
\begin{tabular}[t]{crr}
   \toprule
 & \multicolumn{1}{c}{$\zeta$}
 & events
\\ \midrule \multirow{2}{*}{HH27}
 & $\neq 1\times 10^{-9}$
 & 22280
\\
 & $1 \times 10^{-9}$
 & 222800 \bigstrut[b]
\\ \multirow{2}{*}{HH35}
 & $\neq 1\times 10^{-9}$
 & 10000 \bigstrut[t]
\\
 & $1\times 10^{-9}$
 & 100000 \bigstrut[b]
\\ \multirow{2}{*}{HH50}
 & $\neq 1\times 10^{-9}$
 & 10000 \bigstrut[t]
\\
 & $ 1\times 10^{-9}$
 & 100000 \bigstrut[b]
\\ \multirow{2}{*}{HH65}
 & $\neq 1\times 10^{-9}$
 & 10000 \bigstrut[t]
\\
 & $ 1\times 10^{-9}$
 & 100000 \bigstrut[b]
\\ HH80
 & \multicolumn{1}{c}{all $\zeta$}
 & 10000
\\ \bottomrule
\end{tabular}
\label{tab:signal}
}
\caption{Monte Carlo samples of SM background and signal events used for our analysis.}
\label{tab:signal_and_backgrounds}
\end{table}

\subsection{Event simulation}

All Monte Carlo samples were generated using parton distribution functions given by CTEQ6L1 \cite{Pumplin:2002vw}.
For the simulation of the background we used \software{madgraph} 4.4.44 \cite{Alwall:2007st} interfaced with \software{Pythia} 6.4.22 \cite{Sjostrand:2006za}.

Our simulation of the signal events relied on the following procedure.
First, supersymmetric mass spectra were calculated with a modified version of \software{softsusy} 3.1.5 \cite{Allanach:2009bv} assuming mSUGRA boundary conditions and R-parity conservation.
The latter assumption is justified due to the tiny amount of R-parity breaking in our model.
The \software{softsusy} version was modified in order to produce additionally to the spectrum the R-parity violating neutralino decay width and branching ratios according to Equation~\eqref{Neutralino_Decay_Width}.
The \software{softsusy} mass spectra were fed into \software{sdecay} \cite{Muhlleitner:2004mka} via the \software{madgraph} homepage \cite{madgraph} in order to calculate the decay widths of the SUSY particles (besides the neutralino LSP).
In the next step neutralino decay information was included into the \software{sdecay} output.
The signal process (production of $\widetilde g\widetilde g$, $\widetilde g\widetilde q$, $\widetilde q\widetilde q$ and $\widetilde q \overline{\widetilde q}$) was simulated with \software{madgraph} and then given to \software{pythia} for computation of all subsequent decays according to the \software{sdecay} output as well as for parton showering and hadronization. Table~\ref{tab:signal} shows the signal samples used in our analysis.

The generic detector simulation \software{delphes}, tuned to the CMS detector, was used in order to account for effects of event reconstruction at the detector level.
However, \software{delphes} describes the detector geometry solely in terms of angular variables, i.e. the detector is stretched infinitely in the radial direction.
This approximation is sufficient for most studies involving prompt decays but is untenable in the case of late decaying particles.
We overcome this obstacle by adding vertex information from particles at the
generator level to objects at the detector level. Usually, this information is
provided by the detector simulation. Our procedure is described in detail
in the following section.
We emphasize that a full detector simulation, which includes vertex
reconstruction, needs to be done to improve our analysis.

\subsection{Muon reconstruction process}
\label{sec:physreco}

\begin{figure}
\centering
\psfrag{zinm}{\small$z$ [\unit{m}]}
\psfrag{rinm}{\small$r$ [\unit{m}]}
\psfrag{Muonsystem}{\footnotesize Muon System}
\psfrag{Magnet}{\footnotesize Magnet}
\psfrag{Hadroniccalorimeter}{\footnotesize HCAL}
\psfrag{Electromagneticcal}{\footnotesize ECAL}
\psfrag{Tracker}{\footnotesize Tracker}
\psfrag{Pixel}{\footnotesize Pixel}
\psfrag{eta}{\footnotesize$\eta = 2.4$}
\psfrag{10}{\footnotesize10}
\psfrag{12}{\footnotesize12}
\psfrag{0}{\footnotesize0}
\psfrag{2}{\footnotesize2}
\psfrag{4}{\footnotesize4}
\psfrag{6}{\footnotesize6}
\psfrag{8}{\footnotesize8}
\includegraphics[width=0.7\textwidth]{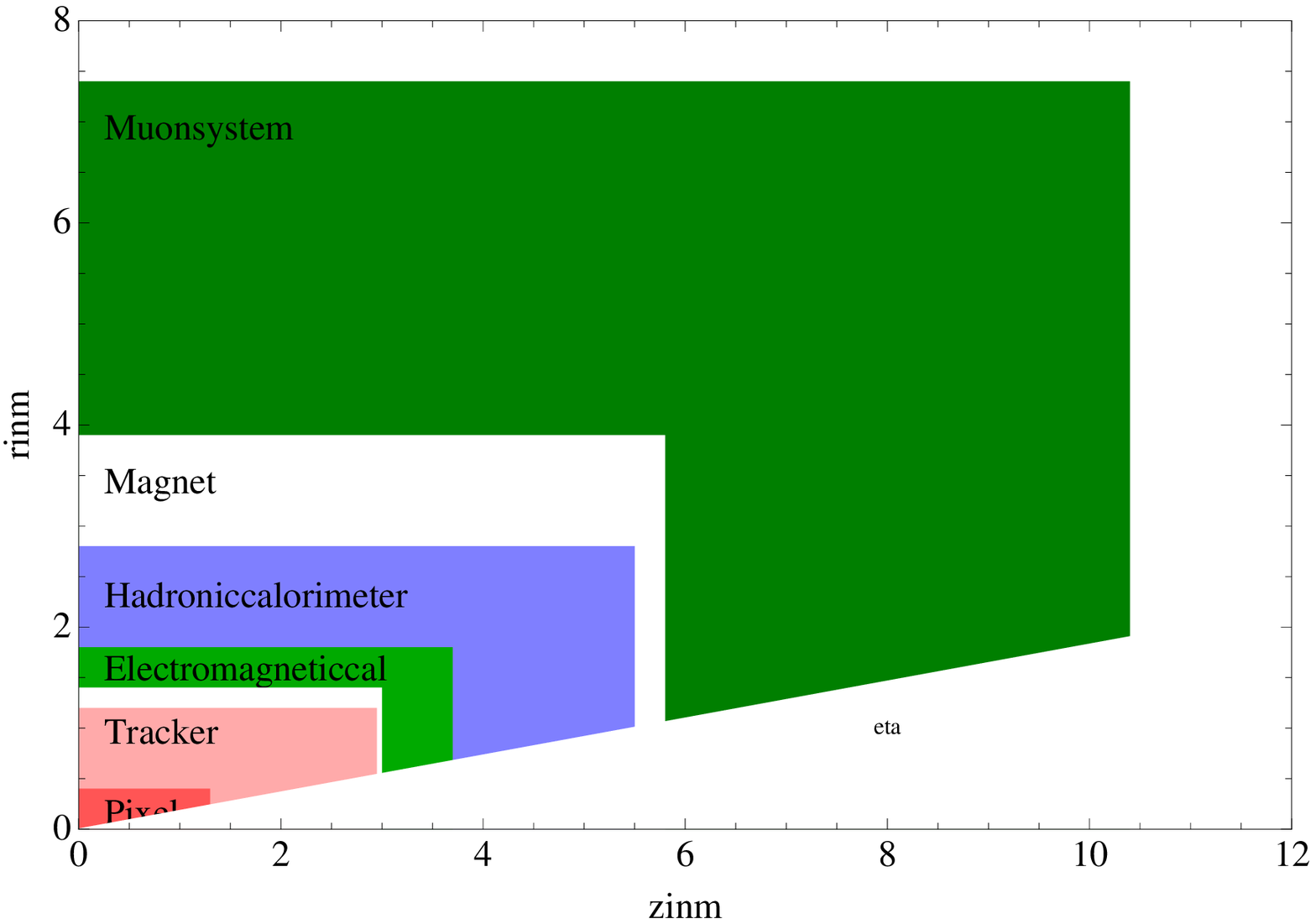}
\caption{Layout of one quarter of the generic detector used for particle identification.}
\label{fig:detector_geometry}
\end{figure}

Particles produced in the late decay of the neutralino will not be properly reconstructed in a real detector if the position of their vertex is beyond or even within the crucial detector component responsible for the respective identification.
For example, an electron produced inside of the electromagnetic calorimeter will leave no track in the tracker and will therefore be identified as a photon
or jet.
In order to simulate the detector response to such events we use a detector geometry in the $(r,z)$ coordinates, which is inspired by the CMS detector at the LHC (see~Fig.~\ref{fig:detector_geometry}).
The angular position of the detector components is given by the CMS tune of \software{delphes}.

In order to be as conservative as possible we only use muon and track objects for the present analysis, since these objects allow a simple simulation of detection efficiency
losses due to the finite size of the detector.
Namely, we assume that a muon can be reconstructed as long as its vertex is in front of the muon chambers, and analogously a track can be reconstructed if it originates approximately
in the first third of the tracker (This region is called pixel detector in Fig.~\ref{fig:detector_geometry}).
For the matching between generator level particles and objects reconstructed by \software{delphes} we use the distance in pseudorapidity $\eta$ and azimuthal angle $\phi$, defined as $\Delta R = \sqrt{(\Delta \phi)^2 + (\Delta \eta)^2}$.

In the following we will call generator level muons, produced by \software{pythia}, \emph{GenMuons}, muons reconstructed initially by \software{delphes}
\emph{muon candidates}, and track objects reconstructed by \software{delphes} \emph{RecoTracks}. Only GenMuons and RecoTracks have the coordinates of their vertex.

First, we perform the following $p_T$ cuts on muon candidates and RecoTracks:
\begin{itemize}
 \item $p_T (\mu) > \unit[20]{GeV}$,
 \item $p_T (\text{Track}) > \unit[15]{GeV}$.
\end{itemize}
These cuts are guided by our SUSY search strategy (c.f.~Section~\ref{sec:neutrsearch}), since we expect that muons coming from $Z$-boson decay have high $p_T$, and a sufficiently high $p_T$ cut can effectively suppress QCD fake leptons.
Furthermore, \software{Delphes} itself reconstructs only muons with $p_T$ above \unit[10]{GeV}.
Additionally, these cuts were optimized in order to get a realistic muon reconstruction efficiency (see~Section~\ref{sec:mueff}).

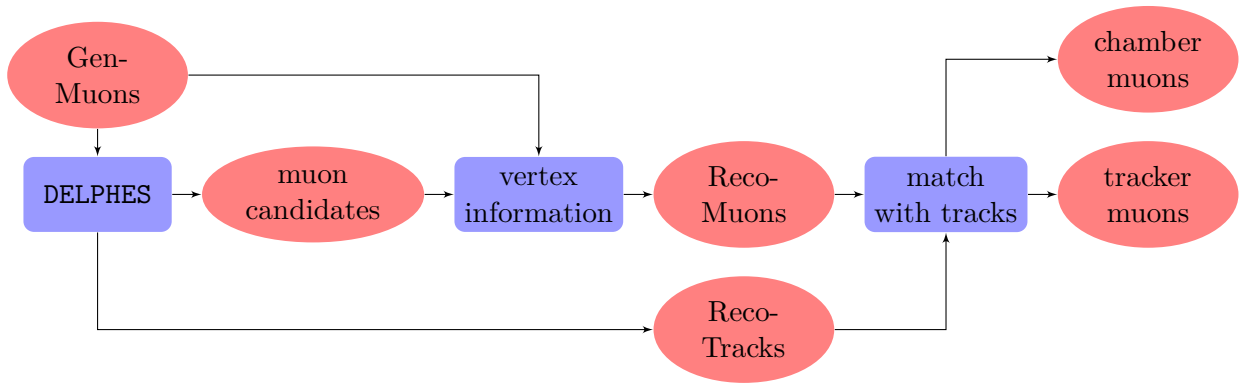
\begin{figure}
\tikzstyle{block} = [rectangle, fill=blue!40, text width=4.4em, text centered, minimum height=6ex, rounded corners]
\tikzstyle{cloud} = [ellipse, fill=red!50, text width=3.7em, text centered%, minimum height=4ex
]
\tikzstyle{line} = [draw, -latex']
\centering
\begin{tikzpicture}[node distance=2.2ex and 1em]
\small
 \node [cloud] (genmuons) {Gen\-Muons};
 \node [block, below =% 4.7ex
 of genmuons] (delphes) {\software{Delphes}};
 \path [line] (genmuons) -- (delphes);
 \node [cloud,text width=4.7em, right = of delphes] (candidates) {muon candidates};
 \path [line] (delphes) -- (candidates);
 \node [block, text width=5.1em, right = of candidates] (vertex) {vertex information};
 \path [line] (genmuons) -| (vertex);
 \path [line] (candidates) -- (vertex);
 \node [cloud, right = of vertex] (recomuons) {Reco\-Muons};
 \path [line] (vertex) -- (recomuons);
 \node [cloud, below = of recomuons] (recotracks) {Reco\-Tracks};
 \path [line] (delphes) |- (recotracks);
 \node [block, text width=4.9em,  right = of recomuons] (trackmatching) {match with tracks};
 \path [line] (recomuons) -- (trackmatching);
 \path [line] (recotracks) -| (trackmatching);
 \node [cloud, right = of trackmatching] (trackermuons) {tracker muons};
 \path [line] (trackmatching) -- (trackermuons);
 \node [cloud, above = of trackermuons] (chambermuons) {chamber muons};
 \path [line] (trackmatching) |- (chambermuons);
 %\node [above = 0ex of candidates] (simulation) {detector simulation};
 %\node [draw=blue!40, rounded corners, fit=(delphes) (candidates) (vertex) (simulation)] {};
\end{tikzpicture}
\caption{Muon reconstruction process.}
\label{fig:flowchart}
\end{figure}

In the second step vertex information is added to the muon candidates by
matching with GenMuons:
\begin{itemize}
 \item A GenMuon is selected for matching with muon candidates if its vertex lies in front of the muon system :
       $r_\mu = \sqrt{x^2 +y^2} < \unit[4000]{mm}$, $\left |z_{\mu}\right | < \unit[6000]{mm}$ (see~Fig.~\ref{fig:detector_geometry}).
 \item The $\Delta R$ distance between each selected GenMuon and all muon candidates is computed.
 \item A GenMuon vertex is added to the muon candidate
closest in $\Delta R$, if $\Delta R < 0.1$ and GenMuon and muon candidate have the same charge.
 \item Muon candidates with added vertex information are called
\emph{RecoMuons}.
\end{itemize}

In the final step, muons with or without signal in the tracker are
distinguished:
\begin{itemize}
 \item A RecoTrack is selected for matching with RecoMuons
       if the track vertex lies in the following range: $r_{T} < \unit[400]{mm}$ , $\left |z_T \right | < \unit[1300]{mm}$.
 \item Each selected RecoTrack is matched with the RecoMuon closest in $\Delta R$, if $\Delta R < 0.1$.
 \item Matched RecoTracks and RecoMuons are called \emph{tracker muons}.
RecoMuons which cannot be matched with RecoTracks are called \emph{chamber muons}. Each RecoMuon is therefore either a tracker muon or a chamber muon.
\end{itemize}

After the reconstruction procedure one is left with two kinds of muon objects: (i) chamber muons which have no track in the tracker and are therefore reconstructed solely by the muon chambers, and (ii) tracker muons which have a track. The muon reconstruction process is depicted in Fig.~\ref{fig:flowchart}. The $\Delta R$ matching condition has been optimized in order to get a realistic muon reconstruction efficiency (see~next section).

\subsection{Muon reconstruction efficiency}
\label{sec:mueff}

\begin{figure}
\centering
\psfrag{Efficiency}{\footnotesize Efficiency}
\psfrag{r in mm}{\footnotesize$r$~[\unit{mm}]}
\psfrag{z in mm}{\footnotesize$|z|$~[\unit{mm}]}
\psfrag{0.1}{\hspace{-0.2em}\scriptsize 0.1}
\psfrag{0.2}{\hspace{-0.2em}\scriptsize 0.2}
\psfrag{0.3}{\hspace{-0.2em}\scriptsize 0.3}
\psfrag{0.4}{\hspace{-0.2em}\scriptsize 0.4}
\psfrag{0.5}{\hspace{-0.2em}\scriptsize 0.5}
\psfrag{0.6}{\hspace{-0.2em}\scriptsize 0.6}
\psfrag{0.7}{\hspace{-0.2em}\scriptsize 0.7}
\psfrag{0.8}{\hspace{-0.2em}\scriptsize 0.8}
\psfrag{0.9}{\hspace{-0.2em}\scriptsize 0.9}
\psfrag{1000}{\scriptsize\hspace{-0.3em}1000}
\psfrag{2000}{\scriptsize\hspace{-0.3em}2000}
\psfrag{3000}{\scriptsize\hspace{-0.3em}3000}
\psfrag{4000}{\scriptsize\hspace{-0.3em}4000}
\psfrag{5000}{\scriptsize\hspace{-0.3em}5000}
\psfrag{6000}{\scriptsize\hspace{-0.3em}6000}
\psfrag{7000}{\scriptsize\hspace{-0.3em}7000}
\psfrag{0}{\scriptsize\hspace{-0.1em}0}
\subfloat{\label{fig:muon_z_eff}\includegraphics[width=0.49\textwidth]{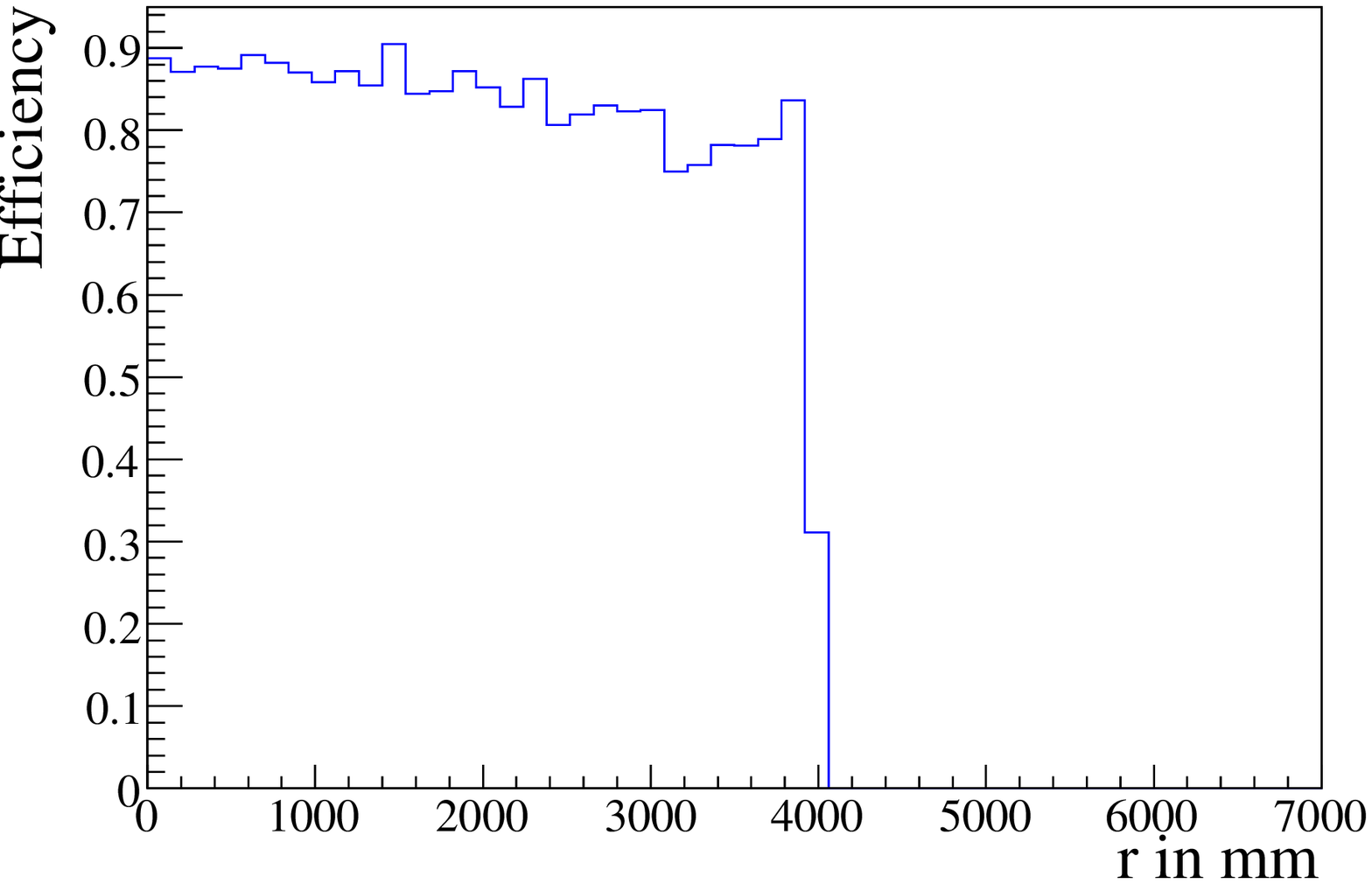}}
\subfloat{\label{fig:muon_r_eff}\includegraphics[width=0.49\textwidth]{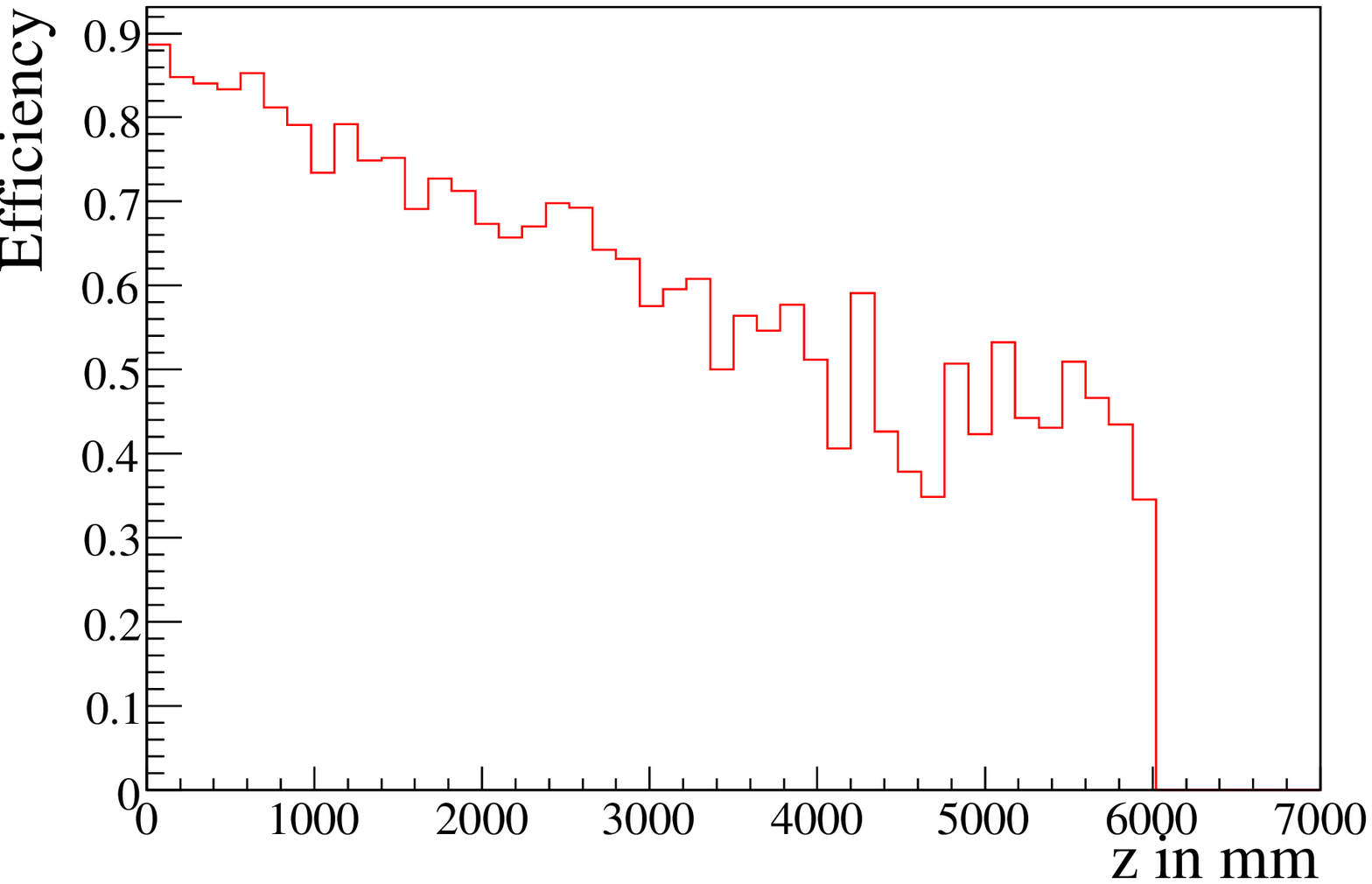}}
\caption{Muon reconstruction efficiency for the benchmark point HH27.}
\label{fig:muon_eff}
\end{figure}
In order to test our method of obtaining physically sensible objects we compute the muon reconstruction efficiency in the following way:
\begin{itemize}
 \item Muons are created as described above.
 \item GenMuons are matched with RecoMuons without any constraints on the position of the GenMuon vertex.
 \item The number of successfully matched objects is compared binwise (in bins of $r$ and $\left|z\right|$) with the number of all GenMuons.
\end{itemize}
The second condition is necessary in order to see whether the assignment between RecoMuons and GenMuons is correct.
Since the matching procedure only relies on angular variables, it is possible that a RecoMuon originally matched with a GenMuon created in front of the muon chamber belongs in fact (i.e. has smaller angular distance) to a GenMuon coming from a decay inside the muon chamber or even outside of the detector.
Such wrong matchings would be seen in the efficiency plot as efficiencies not equal to zero in regions where muons could not be detected by the detector defined above ($r_{\mu} > \unit[4000]{mm}$, $\left|z_{\mu}\right| > \unit[6000]{mm}$).

Fig.~\ref{fig:muon_eff} shows the computed muon efficiency in bins of $r$ and $\left|z\right|$.
As expected one sees a sharp decline in efficiency in the $r$ plot at $r_{\mu} = \unit[4000]{mm}$, where the hard cut applies.
The decline in the $z$ plot is gradually, since physical particles have to fulfill both $r$ and $|z|$ criteria.
The particles originating at small values of $r$ and large values of $|z|$ are not reconstructed due to the limited pseudorapidity coverage of the muon detector.
The efficiency stays at zero beyond $r=\unit[4000]{mm}$ and $\left|z\right| = \unit[6000]{mm}$ as expected, confirming our method of muon reconstruction.
We expect that the computed muon efficiency agrees within \unit[15]{\%} with efficiencies
of present LHC detectors including losses due to muon-jet separation
requirements.

\section{Search for the neutralino decay \texorpdfstring{$\boldsymbol{\chi^0_1 \rightarrow Z \nu}$}{into Z and neutrino}}
\label{sec:neutrsearch}

As described in Section~\ref{sec:search_strategies} our study is focused on the channel $\chi^0_1\rightarrow Z\nu \rightarrow \mu^+ \mu^- \nu$.
This channel possesses certain physical and technical advantages.
On the physical side reliable muon identification is possible already in the early stage of the LHC data taking and one can assume that QCD background can hardly fake two muons at the same time.
Furthermore this signal leads to spectacular events and has no easily identifiable SM background at all, as shown in this section.
Additionally, the muon chamber is the detector component which is farthermost away from the primary vertex and hence one can expect that it will be possible to detect a significant number of clean late time decays even for very small R-parity breaking.
On the technical side, muons seem to be the simplest objects for which a realistic detector response can be modeled within \software{delphes} (see~Section~\ref{sec:physreco}),
due to the limitations of this simulation in the presence of secondary vertices.

The spectacular feature of this signal are opposite sign muon pairs with invariant mass close to the $Z$-boson mass, which have either associated tracks in the tracker with clearly visible secondary vertices or no associated tracks at all.
Such muon pairs cannot be generated by usual SM background as will be shown in the following.
However, a similar signal can arise from cosmic muons traversing the detector.
We could not create a Monte Carlo background sample for cosmic muons, and we simply assume that such background can be suppressed by use of the full timing information of the event:
cosmic muons will first cause a signal in the muon chamber which is closest to the ceiling of the experimental hall followed by a signal in the opposite direction.

An intrinsic background for the presented search are muon pairs from R-parity violating decays, where one muon is coming from the $W$-boson decay while the other muon is coming either from the neutralino decay into the $W$-boson in either of the two branches or from the $W$- or $Z$-boson decay in the second branch (c.f.~Fig.~\ref{fig:decaychain}).
This background can be suppressed if one has access to the corresponding tracks by demanding that both of them originate from the same vertex.
In the case of muons without tracks this background is irreducible. However it belongs itself to the signal one is looking for.

\subsection{Event selection}
\label{sec:H270}

In order to find the signal we now employ a series of simple cuts on the reconstructed objects (muons, tracker muons, and chamber muons).
As we will see, already with an integrated luminosity of only 1 $\unit{fb}^{-1}$ at $\sqrt{s} = \unit[7]{TeV}$ a discovery of the benchmark scenario HH27 with $\zeta = 3\times 10^{-8}$ is possible.

First, we perform a selection cut on the number of muons in the event:
\begin{itemize}
 \item $N(\text{muons}) \ge 2$.
\end{itemize}
We define two event classes depending on the number of tracker muons:
\begin{itemize}
 \item Class 1: the event contains at least two tracker muons $N(\text{tracker muons}) \ge 2$.
 \item Class 2: otherwise.
\end{itemize}
From the description of the signal presented above, we implement additionally two sets of cuts depending on the class of the event.
The cuts for Class 1 events are:
\begin{itemize}
 \item
  All possible invariant masses of opposite sign tracker muons are computed.
  The event passes the cut if at least one invariant mass is in the range of the $Z$-boson mass: $\unit[80]{GeV} < M_{\mu^+\mu^-} < \unit[100]{GeV}$.
  If the event contains more than one appropriate combination of the tracker muons then the muons from the combination with invariant mass closest to the $Z$-boson mass are selected for further analysis.
 \item $d(\text{Vertex}) > \unit[5]{mm}$:
  Each of the tracks associated with the two selected tracker muons should have a vertex which is further than \unit[5]{mm} away from the primary vertex.
  This value is approximately one order of magnitude larger than the current resolution of the inner tracker (c.f.~\cite{Bayatian:2006zz, Aad:2009wy}).
 \item $\Delta d(\text{Vertex})_{ij} < \unit[5]{mm}$:
  The distance between the two track vertices should be less than \unit[5]{mm}.
 \item
  If the event fails one of the cuts it is classified as a Class 2 event.
\end{itemize}
The cuts for Class 2 events are:
\begin{itemize}
 \item $N(\text{chamber muons}) \ge 2$:
  If an event has less than two tracker muons than it should have at least two chamber muons.
 \item
  All possible invariant masses of opposite sign chamber muons are computed.
  An event passes the cut if at least one invariant mass is in the range of the $Z$-boson mass: $\unit[80]{GeV} < M_{\mu^+\mu^-} < \unit[100]{GeV}$.
\end{itemize}
Since each Class 1 event is classified as a Class 2 event if it fails one of the cuts, no signal event is discarded because of the presence of muons with tracks not coming from neutralino decay.

\begin{table}
\centering
\begin{tabular}{rrrrrrr}
   \toprule
 & \multicolumn{4}{c}{background}
 & \multicolumn{2}{c}{signal}
\\ \cmidrule(r){2-5}
 & \multicolumn{1}{c}{$t\bar t$}
 & \multicolumn{1}{c}{$Z$}
 & \multicolumn{1}{l}{di-}
 & \multicolumn{1}{l}{tri-}
 & \multicolumn{2}{c}{$\zeta$}
\\ \cmidrule(r){4-5} \cmidrule(l){6-7}
 &
 &
 & \multicolumn{2}{c}{boson}
 & $3\times10^{-8}$
 & $10^{-9}$
\\ \midrule
   before cuts
 & 200000
 & 700000
 & 150000
 & 40629
 & 22280
 & 220000
\\ $N(\text{muons})\ge2$
 & 3044
 & 9458
 & 2826
 & 1506
 & 2912
 & 4404
\\ \midrule
   Is Class 1
 & 3044
 & 9458
 & 2826
 & 1506
 & 1049
 & 4342
\\ $\unit[80]{GeV}<M_{\mu^+\mu^-}<\unit[100]{GeV}$
 & 337
 & 9118
 & 2418
 & 1051
 & 195
 & 980
\\ $d(\text{Vertex}) > \unit[5]{mm}$
 & 9
 & 0
 & 0
 & 0
 & 49
 & 13
\\ $\Delta d(\text{Vertex})_{ij} < \unit[5]{mm}$
 & 0
 & 0
 & 0
 & 0
 & 36
 & 0
\\ \midrule Is Class 2
 & 3044
 & 9458
 & 2826
 & 1506
 & 2876
 & 4404
\\ $N(\text{chamber muons}) \ge 2$
 & 0
 & 0
 & 0
 & 0
 & 1049
 & 18
\\ $\unit[80]{GeV}<M_{\mu^+\mu^-}<\unit[100]{GeV}$
 & 0
 & 0
 & 0
 & 0
 & 138
 & 2
\\ \midrule
   Total
 & 0
 & 0
 & 0
 & 0
 & 174
 & 2
\\ \bottomrule
\end{tabular}
\caption{Cut flow for HH27 at $\sqrt{s} = \unit[7]{TeV}$. The number of signal events for $\zeta = 3\times 10^{-8}$ ($\zeta = 1\times 10^{-9}$ ) corresponds to an integrated luminosity of $\unit[10]{fb^{-1}}$ ($\approx \unit[100]{fb^{-1}}$).}
\label{tab:cutflow}
\end{table}

Most events will fall into the second class.
The analysis is then very simple and amounts to the search for events with muons without associated track in which the invariant mass of a muon pair lies in the $Z$-boson mass range.
The cut flow is given in Table~\ref{tab:cutflow}.
As expected, no background events survived the cuts, since no standard model process should produce secondary vertices so far away from the primary interaction point.
Although our background estimate has an uncertainty due to the limited statistics, we assume on physical grounds that no background events will pass the cuts if we increase the number of simulated events.
However, the major uncertainty in this study, the number of the background events from cosmic muons, cannot be estimated with the present software.
Therefore a full fledged analysis with full detector simulation which takes into account the cosmic muon background is needed.
In the following we assume that this background can be efficiently suppressed with the full timing information of the event as described in the introduction to Section~\ref{sec:neutrsearch}. Furthermore, we only estimate the systematic uncertainty due to the background and neglect statistical errors and the uncertainty of the muon reconstruction efficiency.

The significance of the signal is computed with the profile likelihood method~\cite{Cowan:2010js} incorporated in the \software{SigCalc} code~\cite{sigcalc}. We assume an integrated luminosity of $\unit[1]{fb^{-1}}$ at $\sqrt{s} = \unit[7]{TeV}$ LHC and a ten times higher Monte Carlo luminosity $\mathcal{L}_{MC} = N_b/\sigma_b = \unit[10]{fb^{-1}}$ for all the background events. At this integrated luminosity 17 signal events and no background events survive the cuts, which corresponds to a significance $Z_{PL} = 9.03$.
Instead, if one makes the pessimistic estimate that 1 background event from the cosmic muons passes the cuts one finds a significance $Z_{PL} = 6.39$. Therefore we conclude that at the benchmark point HH27 with $\zeta = 3\times 10^{-8}$, R-parity breaking neutralino decays can be discovered with the first inverse femtobarn of LHC data.

\begin{figure}
\centering
\psfrag{zinm}{$\left|z\right|$~[\unit{m}]}
\psfrag{rinm}{$r$~[\unit{m}]}
\psfrag{0}{\footnotesize0}
\psfrag{1}{\footnotesize1}
\psfrag{2}{\footnotesize2}
\psfrag{3}{\footnotesize3}
\psfrag{4}{\footnotesize4}
\psfrag{6}{\footnotesize6}
\psfrag{8}{\footnotesize8}
\psfrag{10000}{\footnotesize10000}
\psfrag{10}{\footnotesize10}
\psfrag{12}{\footnotesize12}
\psfrag{0.1}{\footnotesize0.1}
\psfrag{0.2}{\footnotesize}
\psfrag{0.3}{\footnotesize}
\psfrag{0.6}{\footnotesize}
\psfrag{1.}{\footnotesize1}
\psfrag{2.}{\footnotesize}
\psfrag{3.}{\footnotesize}
\psfrag{6.}{\footnotesize}
\psfrag{10.}{\footnotesize10}
\psfrag{20.}{\footnotesize}
\psfrag{30.}{\footnotesize}
\psfrag{60.}{\footnotesize}
\psfrag{100.}{\footnotesize100}
\psfrag{200.}{\footnotesize}
\psfrag{300.}{\footnotesize}
\psfrag{600.}{\footnotesize}
\psfrag{1000.}{\footnotesize1000}
\psfrag{2000.}{\footnotesize}
\psfrag{3000.}{\footnotesize}
\psfrag{6000.}{\footnotesize}
\psfrag{000.}{\footnotesize\hspace{0.3em}000}
\psfrag{9947.}{\footnotesize\colorbox{white}{9947}}
\psfrag{21}{\footnotesize\colorbox{white}{21252}}
\psfrag{252.}{}
\psfrag{26}{\footnotesize\colorbox{white}{26306}}
\psfrag{306.}{}
\psfrag{32}{\footnotesize\colorbox{white}{32203}}
\psfrag{203.}{}
\psfrag{35}{\footnotesize\colorbox{white}{35540}}
\psfrag{540.}{}
\psfrag{41}{\footnotesize\colorbox{white}{41149}}
\psfrag{149.}{}
\psfrag{Muonsystem}{}
\psfrag{Magnet}{}
\psfrag{Hadroniccalorimeter}{}
\psfrag{Electromagneticcalorimeter}{}
\psfrag{Tracker}{}
\psfrag{Pixel}{}
\includegraphics[width=0.9\textwidth]{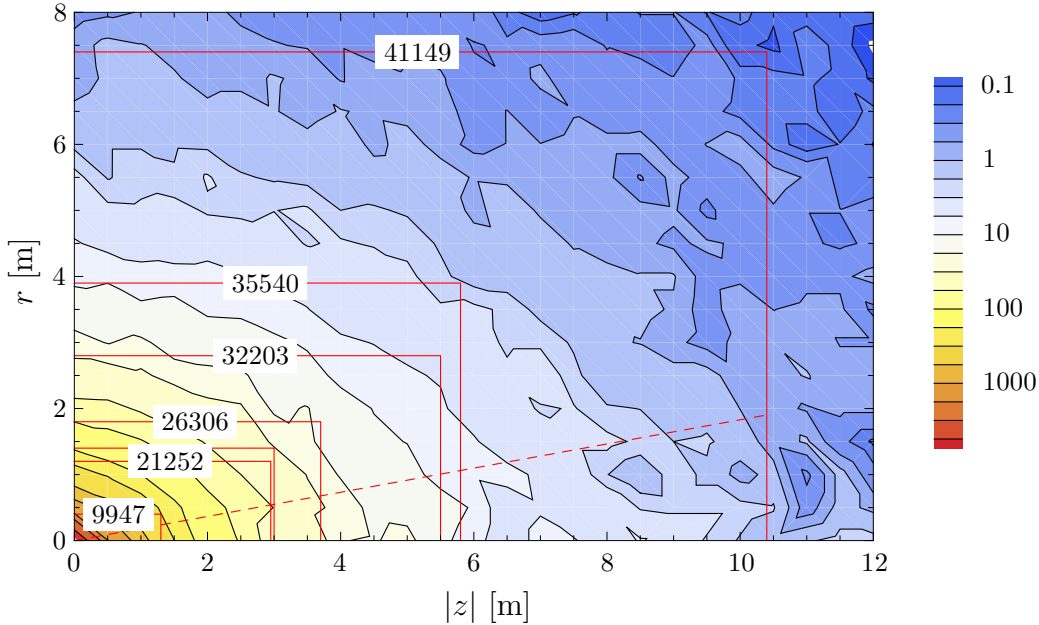}
\caption{Contour plot for the density of neutralino decays inside the detector
per $\unit{m^{-3}}$;
the numbers on the horizontal boundaries of the detector components correspond
to the total number of decays in the enclosed volume;
$m_{\nicefrac{1}{2}} = m_0 = \unit[270]{GeV}$, $\zeta = 3\times10^{-8}$ and
$\mathcal{L} = \unit[10]{fb^{-1}}$.
}
\label{fig:genlevel_decays}
\end{figure}
\begin{figure}
\centering
\psfrag{zinm}{$\left|z\right|$~[\unit{m}]}
\psfrag{rinm}{$r$~[\unit{m}]}
\psfrag{0}{\footnotesize0}
\psfrag{2}{\footnotesize2}
\psfrag{4}{\footnotesize4}
\psfrag{6}{\footnotesize6}
\psfrag{8}{\footnotesize8}
\psfrag{10}{\footnotesize10}
\psfrag{12}{\footnotesize12}
\psfrag{282.}{\footnotesize\colorbox{white}{282}}
\psfrag{236.}{\footnotesize\colorbox{white}{236}}
\psfrag{215.}{\footnotesize\colorbox{white}{215}}
\psfrag{175.}{\footnotesize\colorbox{white}{175}}
\psfrag{139.}{\footnotesize\colorbox{white}{139}}
\psfrag{66.}{\footnotesize\colorbox{white}{66}}
\includegraphics[width=0.7\textwidth]{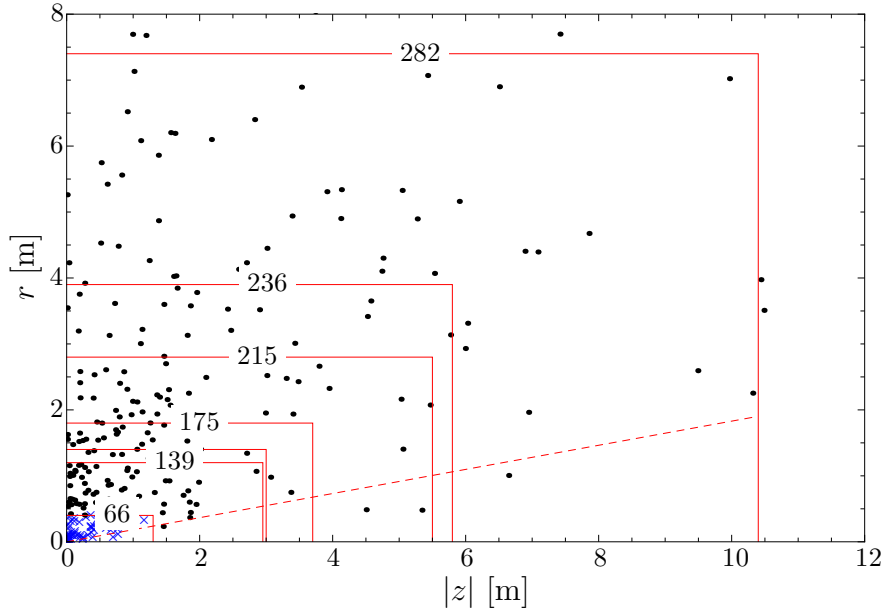}
\caption{Location of secondary vertices for the decays $\chi_1^0 \rightarrow
Z\nu$ with $Z \rightarrow \mu^+\mu^-$ (blue crosses: inside pixel detector,
black dots: outside pixel detector); the numbers on the horizontal boundaries
of the detector components give the number of decays in the enclosed volume;
$m_{\nicefrac{1}{2}} = m_0 = \unit[270]{GeV}$, $\zeta = 3\times10^{-8}$ and
$\mathcal{L} = \unit[10]{fb^{-1}}$.}
\label{fig:decay_density}
\end{figure}

\subsection{Discovery reach at the LHC}
\label{sec:reach}

In the previous section we have studied in detail the benchmark point HH27:
$m_{\nicefrac{1}{2}} = m_0 = \unit[270]{GeV}$, which yields the rather small superparticle masses $m_{\chi_1^0} = \unit[106]{GeV}$, $m_{\tilde g} \simeq \unit[660]{GeV}$ and $m_{\tilde q} \simeq \unit[650]{GeV}$ for the light quark flavors (cf.~Table~\ref{tab:parameter_point_definition}).
From the decay rates given in Section~\ref{sec:neutralino_signatures} and the phase space factors shown in Fig.~\ref{fig:phasespace} one obtains for decay length and branching ratio into $Z$-boson/neutrino final states:
\begin{align}
  c \tau_{\chi_1^0}
 \simeq\unit[4.5]{m} \left( \frac{\zeta}{10^{-8}} \right)^{-2} ,
  \quad
 BR(\chi_1^0 \rightarrow Z\nu)
  \simeq 0.14\ .
\label{BRZ}
\end{align}
Based on the production cross sections listed in Table~\ref{tab:crosssections} an integrated luminosity $\mathcal{L} = \unit[10]{fb^{-1}}$ yields about 22000 events and therefore 44000 NLSPs.

We have studied this benchmark point for two different values of the R-parity breaking parameter:
$\zeta = 3\times10^{-8}$ and $\zeta = 1\times10^{-9}$.
For the larger value of $\zeta$ one has $c\tau_{\chi_1^0}\simeq\unit[50]{cm}$.
Hence, essentially all neutralinos decay inside the detector, most of them close to the origin.
The spacial distribution of secondary vertices is displayed in the contour plot Fig.~\ref{fig:genlevel_decays}.
Using $BR(Z \rightarrow \mu^+\mu^-) \simeq 0.034$ and the branching ratio given in Eq.~\eqref{BRZ}, one concludes that there are about 200 events with a secondary $\chi_1^0$-decay vertex, which contain a $\mu^+\mu^-$ pair with $M_{\mu^+\mu^-} \simeq M_Z$.
This is consistent with the simulation which yields 174 events passing all cuts (cf.~Table~\ref{tab:cutflow}).
The locations of the secondary vertices of these events are shown in Fig.~\ref{fig:decay_density}.

\begin{figure}
\centering
\psfrag{zinm}{$\left|z\right|$~[\unit{m}]}
\psfrag{rinm}{\small$r$~[\unit{m}]}
\psfrag{0}{\footnotesize0}
\psfrag{2}{\footnotesize2}
\psfrag{4}{\footnotesize4}
\psfrag{6}{\footnotesize6}
\psfrag{8}{\footnotesize8}
\psfrag{10}{\footnotesize10}
\psfrag{12}{\footnotesize12}
\psfrag{204.}{\footnotesize\colorbox{white}{204}}
\psfrag{103.}{\footnotesize\colorbox{white}{103}}
\psfrag{71.}{\footnotesize\colorbox{white}{71}}
\psfrag{41.}{\footnotesize\colorbox{white}{41}}
\psfrag{26.}{\footnotesize\colorbox{white}{26}}
\psfrag{8.}{\footnotesize\colorbox{white}8}
\includegraphics[width=0.7\textwidth]{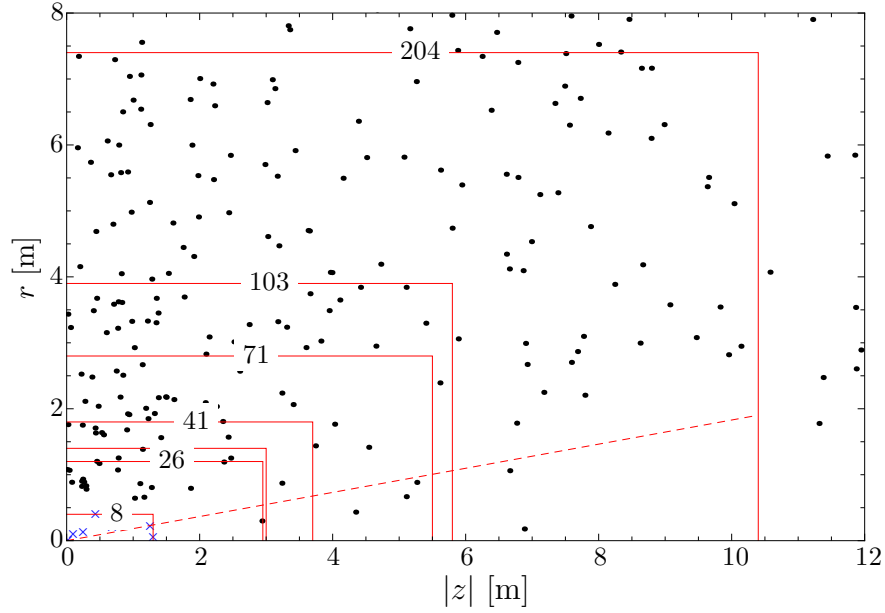}
\caption{Location of all neutralino decays inside of the detector
(blue crosses: decays inside pixel detector; black dots:
decays outside pixel detector); the numbers on the horizontal boundaries of
the detector components correspond to the total number of decays in the
enclosed volume; $m_{\nicefrac{1}{2}} = m_0 = \unit[270]{GeV}$, $\zeta = 1\times10^{-9}$
and $\mathcal{L} = \unit[10]{fb^{-1}}$.}
\label{fig:HH27_z1e9}
\end{figure}

For the smaller value of the R-parity breaking parameter, $\zeta = 1\times10^{-9}$, the decay length increases to $c\tau_{\chi_1^0}\simeq\unit[450]{m}$.
Now most neutralino NLSPs decay outside the detector.
This is apparent from Fig.~\ref{fig:HH27_z1e9} where the total number of decays in the different subvolumina of the detector are given.
Compared to $\zeta = 3\times10^{-8}$, the number of decays inside the detector is smaller by a factor $\sim 200$, which roughly corresponds to the ratio of the decay lengths, as suggested in \cite{Ishiwata:2008tp}.

According to the simulation described in the previous section, for $\zeta = 1\times10^{-9}$ an integrated luminosity of $\unit[100]{fb^{-1}}$ is needed to obtain 2 decays $\chi_1^0 \rightarrow Z\nu \rightarrow \mu^+\mu^-\nu$, inside the detector.
This number is very small compared to the total number of about 2000 decays in the detector, which is a consequence of the tiny branching ratio into the chosen specific final state.
It is likely that a substantially larger fraction of the events can be used in the search for a decaying neutralino.
In \cite{Ishiwata:2008tp} it has been argued that already 10 $\chi_1^0$ decays inside the detector may be sufficient for the discovery of a decaying NLSP, which would require an integrated luminosity of only $\unit[1]{fb^{-1}}$.
It remains to be seen whether for events with a secondary vertex and jets, signal and background can be sufficiently well separated.

\begin{figure}
\centering
\psfrag{zinm}{$\left|z\right|$~[\unit{m}]}
\psfrag{rinm}{\small$r$~[\unit{m}]}
\psfrag{0}{\footnotesize0}
\psfrag{1}{\footnotesize1}
\psfrag{2}{\footnotesize2}
\psfrag{4}{\footnotesize4}
\psfrag{6}{\footnotesize6}
\psfrag{8}{\footnotesize8}
\psfrag{10}{\footnotesize10}
\psfrag{12}{\footnotesize12}
\psfrag{1.}{\footnotesize\colorbox{white}{1}}
\psfrag{2.}{\footnotesize\colorbox{white}{2}}
\psfrag{6.}{\footnotesize\colorbox{white}{6}}
\psfrag{8.}{\footnotesize\colorbox{white}{8}}
\psfrag{12.}{\footnotesize\colorbox{white}{12}}
\includegraphics[width=0.7\textwidth]{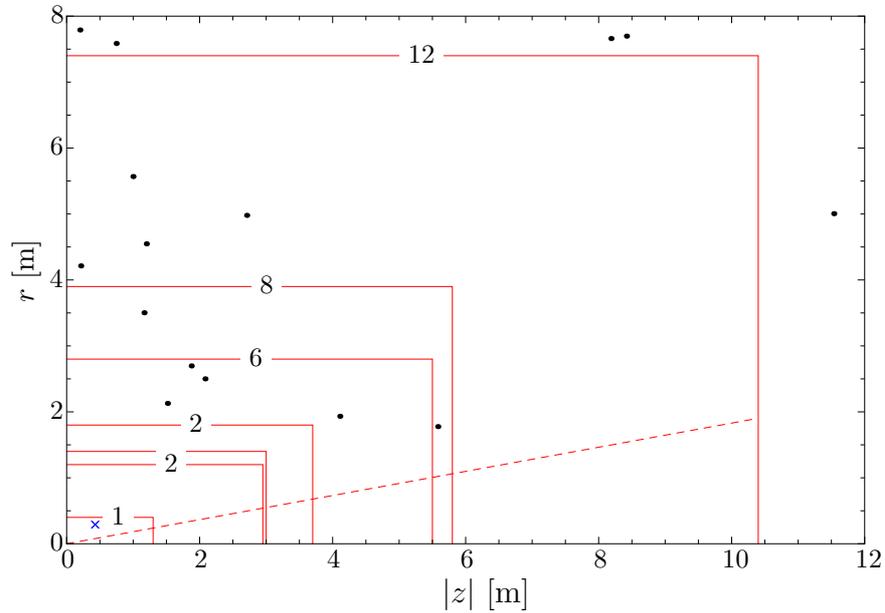}
\caption{Location of all neutralino decays inside of the detector
(blue cross: decay inside pixel detector; black dots:
decays outside pixel detector); the numbers on the horizontal boundaries of
the detector components correspond to the total number of decays in the
enclosed volume; $m_{\nicefrac{1}{2}} = m_0 = \unit[500]{GeV}$, $\zeta = 1\times10^{-9}$
and $\mathcal{L} = \unit[10]{fb^{-1}}$.}
\label{fig:HH50_z1e9}
\end{figure}

Let us now consider the benchmark point HH50:
$m_{\nicefrac{1}{2}} = m_0 =\unit[500]{GeV}$, which implies the heavier superparticle
masses $m_{\chi_1^0} =\unit[206]{GeV}$ and
$m_{\tilde g} \simeq m_{\tilde q} \simeq \unit[1200]{GeV}$ for
the light quark flavours (cf.~Table~\ref{tab:parameter_point_definition}).
The phase space suppression is now negligible, $f_W \simeq f_Z \simeq 1$, and
one obtains for decay length and branching ratio into $Z$-boson/neutrino final
states:
\begin{align}
   c \tau_{\chi_1^0}
 \simeq \unit[0.8]{m} \left( \frac{\zeta}{10^{-8}} \right)^{-2} ,
 \quad
  BR(\chi_1^0 \rightarrow Z\nu)
 \simeq
  0.33\ .
\label{BRZ2}
\end{align}
The total production cross section for these heavier gluino/squark pairs is
about two orders of magnitude smaller (cf.~Table~\ref{tab:crosssections}),
and therefore an integrated luminosity $\mathcal{L} = \unit[10]{fb^{-1}}$
only yields 460 NLSPs.

We have studied this benchmark point again for the two different values of
the R-parity breaking parameter $\zeta = 3\times10^{-8}$ and
$\zeta = 1\times10^{-9}$.
For the larger value of $\zeta$ one has $c\tau_{\chi_1^0}\simeq\unit[10]{cm}$,
and essentially all neutralinos decay inside the detector.
The branching ratio into the considered final state is now somewhat larger,
$BR(\chi_1^0 \rightarrow Z\nu \rightarrow \mu^+\mu^-\nu) \simeq 0.01$, so that
one expects about 4 events with this final state, which is consistent with our
simulation. Hence, for this larger value of the R-parity breaking parameter
and this benchmark point, the discovery of a decaying NLSP appears feasible
already in the early phase of the LHC.

For $\zeta = 1\times10^{-9}$, the decay length is
$c\tau_{\chi_1^0}\simeq\unit[80]{m}$ and most neutralino NLSPs decay outside
the detector. The spacial distribution of secondary vertices inside the
detector, in total 12 for $\unit[10]{fb^{-1}}$, is shown in
Fig.~\ref{fig:HH50_z1e9}. Due to the \unit[1]{\%} branching ratio into the
$Z\nu \rightarrow \mu^+\mu^-\nu$ final state one then estimates that
$\unit[1000]{fb^{-1}}$ will be needed for a discovery, which is consistent
with our simulation.

\begin{figure}
\centering
\psfrag{zeta}{$\zeta$}
\psfrag{Integrated Luminosity}{Integrated Luminosity [$\unit{fb^{-1}}$]}
\psfrag{discoveryreach}{\hspace{-2.5em}Discovery Reach~[$\unit{fb^{-1}}$]}
\psfrag{-}{\footnotesize}
\psfrag{0}{\footnotesize}
\psfrag{1}{\footnotesize1}
\psfrag{2}{\footnotesize2}
\psfrag{3}{\footnotesize3}
\psfrag{4}{\footnotesize4}
\psfrag{8}{\footnotesize\hspace{-1.3em}$10^{-8}$}
\psfrag{9}{\footnotesize\hspace{-1.3em}$10^{-9}$}
\psfrag{10}{\footnotesize\hspace{-1.3em}$10^{-10}$}
\psfrag{-8}{\footnotesize\hspace{-0.8em}$10^{-8}$}
\psfrag{-9}{\footnotesize\hspace{-0.8em}$10^{-9}$}
\psfrag{-10}{\footnotesize\hspace{-0.6em}$10^{-10}$}
\psfrag{10}{\footnotesize\hspace{-0.5em}10}
\psfrag{100}{\footnotesize\hspace{-0.5em}100}
\psfrag{1000}{\footnotesize\hspace{-0.6em}1000}
\psfrag{000}{\footnotesize\hspace{-0.2em}000}
\psfrag{HH27}{\footnotesize HH27}
\psfrag{HH35}{\footnotesize HH35}
\psfrag{HH50}{\footnotesize HH50}
\psfrag{HH65}{\footnotesize HH65}
\psfrag{HH80}{\footnotesize HH80}
\includegraphics[width=0.78\textwidth]{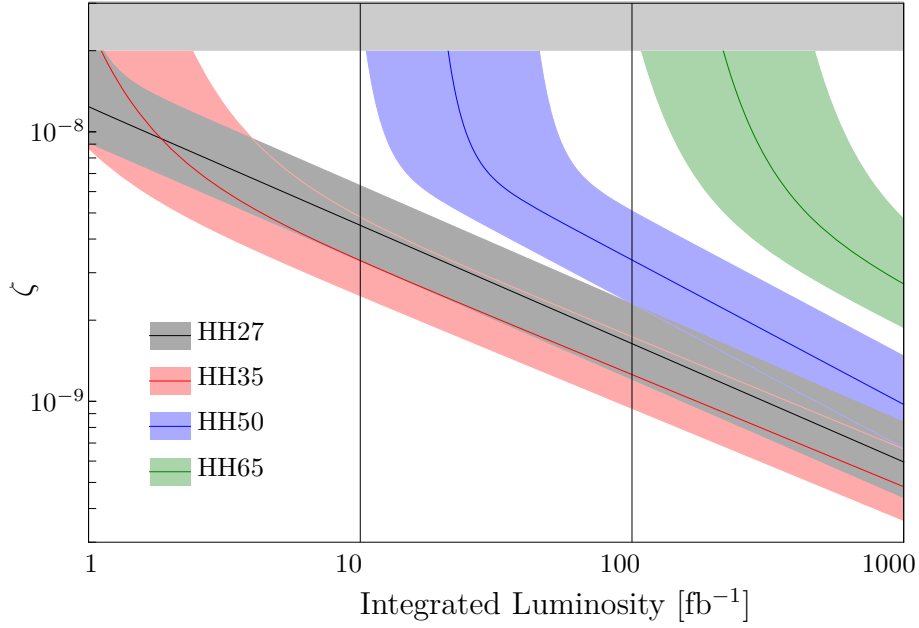}
\caption{$5\sigma$ discovery reach in $\zeta$ for quasi-stable neutralino NLSPs via the
decays $\chi_1^0 \rightarrow Z\nu$ with $Z \rightarrow \mu^+\mu^-$.
The different bench mark points correspond to gluino and squark masses between
$\unit[650]{GeV}$ and $\unit[1800]{GeV}$; the bands represent
different assumptions about the background (see text).}
\label{fig:discovery_reach}
\end{figure}

In Fig.~\ref{fig:discovery_reach} we have summarized the results of our simulations
for the decay chain $\chi_1^0 \rightarrow Z\nu$ with $Z \rightarrow \mu^+\mu^-$.
The benchmark points HH27--HH80 correspond to gluino and squark masses ranging
from $\unit[650]{GeV}$ to $\unit[1800]{GeV}$
(cf.~Table~\ref{tab:parameter_point_definition}).
The bands reflect the
different number of events required for a $5\sigma$ discovery depending on the
simulated background. The central value corresponds to 6 signal events (with
luminosity $\mathcal L$) with no background events for a simulated luminosity of
$10\times\mathcal{L}$; the lower (upper) boundary represents 3 (13) signal events
(with luminosity $\mathcal L$) with no (1) background event for a simulated
luminosity of $100\times\mathcal{L}$ ($10\times\mathcal{L}$).
We conclude that with
$\unit[10]{fb^{-1}}$ a $5\sigma$ discovery of a quasi-stable neutralino is
possible for squark and gluino masses of $\unit[830]{GeV}$ (cf.~HH35) and
an R-parity breaking parameter $\zeta = 3\times10^{-9}$, which is one order
of magnitude smaller than the present astrophysical bound \cite{Bobrovskyi:2010ps,Vertongen:2011mu}.

\begin{figure}
\centering
\psfrag{zeta}{$\zeta$}
\psfrag{Integrated Luminosity}{Integrated Luminosity [$\unit{fb^{-1}}$]}
\psfrag{discoveryreach}{\hspace{-2.5em}Discovery Reach~[$\unit{fb^{-1}}$]}
\psfrag{-}{\footnotesize-}
\psfrag{0}{\footnotesize}
\psfrag{1}{\footnotesize1}
\psfrag{2}{\footnotesize2}
\psfrag{3}{\footnotesize3}
\psfrag{4}{\footnotesize4}
\psfrag{5}{\footnotesize5}
\psfrag{8}{\footnotesize\hspace{-1.3em}$10^{-8}$}
\psfrag{9}{\footnotesize\hspace{-1.3em}$10^{-9}$}
\psfrag{10}{\footnotesize\hspace{-1.3em}$10^{-10}$}
\psfrag{-8}{\footnotesize\hspace{-0.8em}$10^{-8}$}
\psfrag{-9}{\footnotesize\hspace{-0.8em}$10^{-9}$}
\psfrag{-10}{\footnotesize\hspace{-0.6em}$10^{-10}$}
\psfrag{8}{\footnotesize8}
\psfrag{9}{\footnotesize9}
\psfrag{10}{\footnotesize10}
\psfrag{0.01}{\footnotesize\hspace{-0.5em}0.01}
\psfrag{0.1}{\footnotesize\hspace{-0.5em}0.1}
\psfrag{1.}{\footnotesize\hspace{-0.5em}1}
\psfrag{10.}{\footnotesize\hspace{-0.5em}10}
\psfrag{100.}{\footnotesize\hspace{-0.5em}100}
\psfrag{1000.}{\footnotesize\hspace{-0.6em}1000}
\psfrag{10}{\footnotesize10}
\psfrag{100}{\footnotesize100}
\psfrag{1000}{\footnotesize1000}
\psfrag{HH27}{\footnotesize HH27}
\psfrag{HH35}{\footnotesize HH35}
\psfrag{HH50}{\footnotesize HH50}
\psfrag{HH65}{\footnotesize HH65}
\psfrag{HH80}{\footnotesize HH80}
\includegraphics[width=0.78\textwidth]{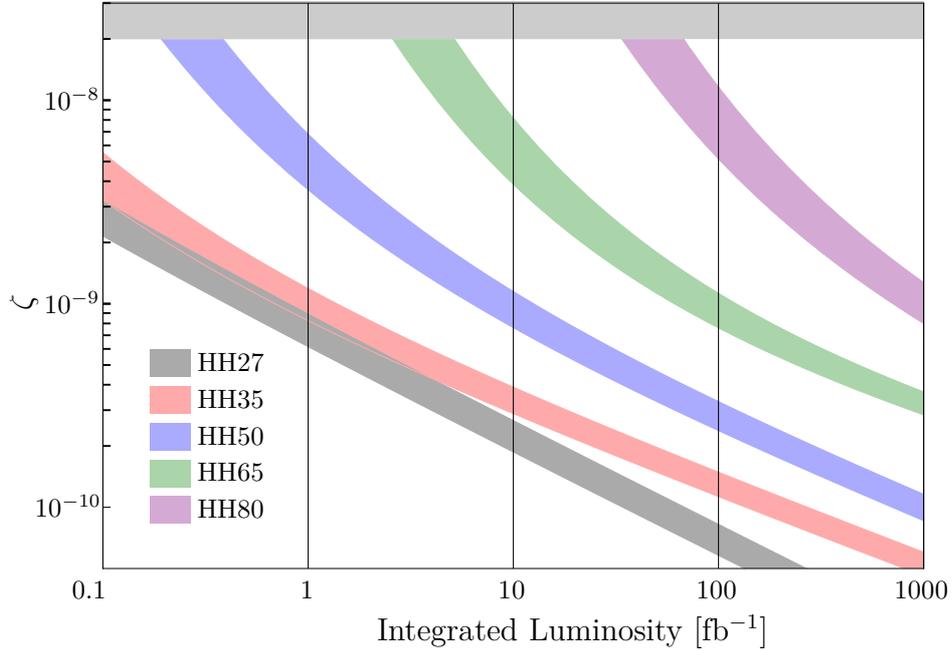}
\caption{Estimate of the $5 \sigma$ discovery reach in $\zeta$ for quasi-stable neutralino NLSPs at the LHC; the lower (upper) boundary of the bands
corresponds to 10 (20) decays inside the detector.
The different bench mark points correspond to gluino and squark masses between $\unit[650]{GeV}$ and $\unit[1800]{GeV}$.
}
\label{fig:optimistic_discovery_reach}
\end{figure}

We expect that the sensitivity in the parameter $\zeta$ can be significantly
improved if also neutralino decays with jets are taken into account.
Fig.~\ref{fig:optimistic_discovery_reach} represents an estimate of the discovery reach for quasi-stable neutralino NLSPs at the LHC, assuming 10--20 decays inside the detector (cf.~\cite{Ishiwata:2008tp}).
The parameter space, which can be probed, is now significantly extended.
As an example, with $\unit[10]{fb^{-1}}$ and squark and gluino masses of \unit[830]{GeV} (cf.~HH35), one is now sensitive to $\zeta = 3\times10^{-10}$, which lies two orders of magnitude below the present astrophysical bound.
Correspondingly, for heavier gluinos and squarks, $m_{\tilde g} \simeq m_{\tilde q} \simeq \unit[1480]{GeV}$ (cf.~HH65), one can probe values of the R-parity breaking parameter down to $\zeta = 3\times10^{-9}$.


\section{Summary and conclusion}
\label{sec:summary}

We have studied the decays of quasi-stable neutralinos in supersymmetric extensions of the Standard Model with small R-parity and lepton number violating couplings which are consistent with primordial nucleosynthesis, thermal leptogenesis and gravitino dark matter.
As representative examples with neutralino NLSP we have considered five benchmark points, HH27--HH80, with gluino and squark masses ranging from $\unit[650]{GeV}$ to $\unit[1800]{GeV}$.

For the considered benchmark points the neutralino is bino-like, and the R-parity breaking parameter $\zeta$, which governs the neutralino decays $\chi_1^0 \rightarrow W^\pm l^\mp$ and $\chi_1^0 \rightarrow Z \nu$, is directly related to the partial gravitino decay width into photon and neutrino.

The main goal of the present work has been to determine the range of the parameter $\zeta$ which can be probed at the LHC, for varying superparticle masses.
As a conservative starting point, we have focused on events with a clean signature:
cascade processes with jets where one of the produced neutralino NLSPs decays into $Z$-boson and neutrino, with a subsequent decay of the $Z$-boson into a muon pair.

In Section~\ref{sec:signatures_and_reach} we have studied the qualitative signatures of these events,
the $\beta\gamma$ distribution of the produced NLSPs,
the $\slashed p_T$-spectrum and the number of leptons in the final state.
A detailed simulation of signal and background events for the generic detector \software{DELPHES} (with CMS tune) has been described in Section~\ref{sec:signal_and_background},
with emphasis on the reconstruction of muons.
The crucial new
element of our analysis has been the implementation of the finite NLSP decay
length.

The numerical results of our simulations have been summarized in Section~\ref{sec:neutrsearch}.
It is very instructive to look at the spacial distribution of decay vertices which have been given for different benchmark points and different values of $\zeta$.
One can clearly see that the sensitivity extends from decay lengths $\mathcal{O}(\unit[50]{cm})$, with NLSP decays mostly inside the detector, to values $\mathcal{O}(\unit[500]{m})$ where almost all NLSPs decay outside the detector.

The results for the discovery reach for quasi-stable neutralino NLSPs roughly agree with the simple estimates which one obtains from the branching ratios into the $Z(\mu^+\mu^-)\nu$ final state together with the assumption that these events are background free.
It is remarkable that already with $\unit[10]{fb^{-1}}$ a $5\sigma$ discovery is possible for squark and gluino masses of \unit[830]{GeV} and an R-parity breaking parameter $\zeta = 3\times10^{-9}$, which is one order of magnitude smaller than the present astrophysical bound.

It is likely that the severe restriction of our simulation to $Z(\mu^+\mu^-)\nu$ final states can be relaxed and that a much larger fraction of events can be used for the analysis.
Assuming optimistically that 10 decays inside the detector are sufficient for a discovery, one would be sensitive to values of $\zeta$ down to $3\times10^{-10}$ for squark and gluino masses of \unit[830]{GeV} and $\unit[10]{fb^{-1}}$.
Alternatively, for gluino and squark masses of \unit[1480]{GeV} one could probe values of the R-parity breaking parameter down to $3\times10^{-9}$.
For the same quark masses a luminosity of $\unit[100]{fb^{-1}}$ would improve the sensitivity in $\zeta$ by a factor of three.
We conclude that for gluino and squark masses accessible at the LHC values of the R-parity breaking parameter can be probed which are far below the present upper bounds obtained from astrophysics and cosmology.

\subsection*{Acknowledgments}

The authors thank F.~Br\"ummer, H.~Schettler, C.~Sander,
G.~Vertongen and especially P.~Schleper for helpful discussions and
comments on the manuscript. This work was supported by the Excellence
Initiative Hamburg ``Connecting Particles with the Cosmos''.

\newpage
\bibliographystyle{h-physrev}
\bibliography{reference}

\begin{thebibliography}{10}

\bibitem{Hall:1983id}
L.~J. Hall and M.~Suzuki,
\newblock Nucl. Phys. {\bf B231}, 419 (1984).

\bibitem{Ross:1984yg}
G.~G. Ross and J.~W.~F. Valle,
\newblock Phys. Lett. {\bf B151}, 375 (1985).

\bibitem{Ellis:1984gi}
J.~R. Ellis, G.~Gelmini, C.~Jarlskog, G.~G. Ross, and J.~W.~F. Valle,
\newblock Phys. Lett. {\bf B150}, 142 (1985).

\bibitem{Allanach:2003eb}
B.~C. Allanach, A.~Dedes, and H.~K. Dreiner,
\newblock Phys. Rev. {\bf D69}, 115002 (2004), hep-ph/0309196.

\bibitem{Barbier:2004ez}
R.~Barbier {\em et~al.},
\newblock Phys. Rept. {\bf 420}, 1 (2005), hep-ph/0406039.

\bibitem{deCampos:2007bn}
F.~{de Campos} {\em et~al.},
\newblock JHEP {\bf 05}, 048 (2008), 0712.2156.

\bibitem{Hirsch:2005ag}
M.~Hirsch, W.~Porod, and D.~Restrepo,
\newblock JHEP {\bf 03}, 062 (2005), hep-ph/0503059.

\bibitem{Takayama:2000uz}
F.~Takayama and M.~Yamaguchi,
\newblock Phys. Lett. {\bf B485}, 388 (2000), hep-ph/0005214.

\bibitem{Buchmuller:2007ui}
W.~Buchm{\"u}ller, L.~Covi, K.~Hamaguchi, A.~Ibarra, and T.~Yanagida,
\newblock JHEP {\bf 03}, 037 (2007), hep-ph/0702184.

\bibitem{Bobrovskyi:2010ps}
S.~Bobrovskyi, W.~Buchm{\"u}ller, J.~Hajer, and J.~Schmidt,
\newblock JHEP {\bf 10}, 061 (2010), 1007.5007.

\bibitem{Lola:2007rw}
S.~Lola, P.~Osland, and A.~R. Raklev,
\newblock Phys. Lett. {\bf B656}, 83 (2007), 0707.2510.

\bibitem{Abdo:2010nc}
Fermi-LAT Collaboration, A.~A. Abdo {\em et~al.},
\newblock Phys. Rev. Lett. {\bf 104}, 091302 (2010), 1001.4836.

\bibitem{Abdo:2010nz}
Fermi-LAT Collaboration, A.~A. Abdo {\em et~al.},
\newblock Phys. Rev. Lett. {\bf 104}, 101101 (2010), 1002.3603.

\bibitem{Vertongen:2011mu}
G.~Vertongen and C.~Weniger,
\newblock JCAP {\bf 1105}, 027 (2011), 1101.2610.

\bibitem{Ishiwata:2008tp}
K.~Ishiwata, T.~Ito, and T.~Moroi,
\newblock Phys. Lett. {\bf B669}, 28 (2008), 0807.0975.

\bibitem{Asai:2009ka}
S.~Asai, K.~Hamaguchi, and S.~Shirai,
\newblock Phys. Rev. Lett. {\bf 103}, 141803 (2009), 0902.3754.

\bibitem{Meade:2010ji}
P.~Meade, M.~Reece, and D.~Shih,
\newblock JHEP {\bf 10}, 067 (2010), 1006.4575.

\bibitem{Desch:2010gi}
K.~Desch, S.~Fleischmann, P.~Wienemann, H.~K. Dreiner, and S.~Grab,
\newblock Phys. Rev. {\bf D83}, 015013 (2011), 1008.1580.

\bibitem{Bomark:2011ye}
N.-E. Bomark, D.~Choudhury, S.~Lola, and P.~Osland,
\newblock (2011), 1105.4022.

\bibitem{Ovyn:2009tx}
S.~Ovyn, X.~Rouby, and V.~Lemaitre,
\newblock (2009), 0903.2225.

\bibitem{Mukhopadhyaya:1998xj}
B.~Mukhopadhyaya, S.~Roy, and F.~Vissani,
\newblock Phys.Lett. {\bf B443}, 191 (1998), hep-ph/9808265.

\bibitem{Aad:2011xm}
ATLAS Collaboration, G.~Aad {\em et~al.},
\newblock (2011), 1103.6214.

\bibitem{Chatrchyan:2011bz}
CMS Collaboration, S.~Chatrchyan {\em et~al.},
\newblock (2011), 1103.1348.

\bibitem{Chatrchyan:2011wb}
CMS Collaboration, S.~Chatrchyan {\em et~al.},
\newblock (2011), 1104.3168.

\bibitem{Nakamura:2010zzi}
Particle Data Group, K.~Nakamura {\em et~al.},
\newblock J. Phys. {\bf G37}, 075021 (2010).

\bibitem{Ball:2007zza}
CMS Collaboration, G.~Bayatian {\em et~al.},
\newblock J.Phys.G {\bf G34}, 995 (2007).

\bibitem{Beenakker:1996ed}
W.~Beenakker, R.~Hopker, and M.~Spira,
\newblock (1996), hep-ph/9611232.

\bibitem{Aad:2009wy}
ATLAS Collaboration, G.~Aad {\em et~al.},
\newblock (2009), 0901.0512.

\bibitem{Kidonakis:2010bb}
N.~Kidonakis,
\newblock PoS {\bf ICHEP2010}, 059 (2010), 1008.2460.

\bibitem{Binoth:2008kt}
T.~Binoth, G.~Ossola, C.~Papadopoulos, and R.~Pittau,
\newblock JHEP {\bf 0806}, 082 (2008), 0804.0350.

\bibitem{Campbell:2011bn}
J.~M. Campbell, R.~Ellis, and C.~Williams,
\newblock (2011), 1105.0020.

\bibitem{Gavin:2010az}
R.~Gavin, Y.~Li, F.~Petriello, and S.~Quackenbush,
\newblock (2010), 1011.3540.

\bibitem{Pumplin:2002vw}
J.~Pumplin {\em et~al.},
\newblock JHEP {\bf 07}, 012 (2002), hep-ph/0201195.

\bibitem{Alwall:2007st}
J.~Alwall {\em et~al.},
\newblock JHEP {\bf 09}, 028 (2007), 0706.2334.

\bibitem{Sjostrand:2006za}
T.~Sjostrand, S.~Mrenna, and P.~Z. Skands,
\newblock JHEP {\bf 05}, 026 (2006), hep-ph/0603175.

\bibitem{Allanach:2009bv}
B.~C. Allanach and M.~A. Bernhardt,
\newblock Comput. Phys. Commun. {\bf 181}, 232 (2010), 0903.1805.

\bibitem{Muhlleitner:2004mka}
M.~Muhlleitner,
\newblock Acta Phys. Polon. {\bf B35}, 2753 (2004), hep-ph/0409200.

\bibitem{madgraph}
J.~Alwall {\em et~al.},
\newblock Madgraph: \url{http://madgraph.hep.uiuc.edu/}.

\bibitem{Bayatian:2006zz}
CMS Collaboration, G.~Bayatian {\em et~al.},
\newblock (2006).

\bibitem{Cowan:2010js}
G.~Cowan, K.~Cranmer, E.~Gross, and O.~Vitells,
\newblock Eur.Phys.J. {\bf C71}, 1554 (2011), 1007.1727.

\bibitem{sigcalc}
G.~Cowan,
\newblock Sigcalc: \url{http://www.pp.rhul.ac.uk/~cowan/stat/SigCalc/}.

\end{thebibliography}

\end{document}